\documentclass[%
 reprint,
 floatfix,   %
superscriptaddress,
 amsmath,amssymb,
 aps,
 prx,
]{revtex4-2}
\usepackage{comment}
\usepackage{graphicx}
\usepackage{dcolumn}
\usepackage{bm}
\usepackage[mathlines]{lineno}
\usepackage{siunitx}
\usepackage{ulem}
\usepackage{todonotes}
\usepackage{physics}
\usepackage{float}
\usepackage{placeins}   %
\usepackage{booktabs}
\usepackage{afterpage}
\usepackage{xcolor}
\usepackage[version=4]{mhchem}
\usepackage{mwe} %
\usepackage{tikz}
\usetikzlibrary{positioning, arrows.meta, fit, calc}
\definecolor{heraAgent}{RGB}{225, 235, 250} %
\definecolor{heraLoop}{RGB}{180, 80, 30}    %
\definecolor{heraGroup}{RGB}{30, 110, 190}   %
\definecolor{lightgray}{gray}{0.6}
\definecolor{medgray}{gray}{0.4}
\definecolor{mRed}{RGB}{230, 0, 50}
\colorlet{newtextColor}{mRed}
\usepackage{blindtext}
\usepackage{hyperref}
\hypersetup{
colorlinks=true,
urlcolor= blue,
citecolor=blue,
linkcolor= blue,
}
\usepackage{etoolbox}
\definecolor{variantA}{HTML}{0B6E8C}
\definecolor{variantB}{HTML}{B45309}
\definecolor{variantC}{HTML}{C2185B}
\newcommand{\markuplevel}{clean}
\newif\ifshowmarkup          %
\newif\ifshowauthornotes     %
\newif\ifshowreviewernotes   %
\ifdefstring{\markuplevel}{all}{\showmarkuptrue\showauthornotestrue\showreviewernotestrue}{}
\ifdefstring{\markuplevel}{author}{\showmarkuptrue\showauthornotestrue\showreviewernotesfalse}{}
\ifdefstring{\markuplevel}{red}{\showmarkuptrue\showauthornotesfalse\showreviewernotesfalse}{}
\ifdefstring{\markuplevel}{clean}{\showmarkupfalse\showauthornotesfalse\showreviewernotesfalse}{}

\ifshowmarkup
    
    \newcommand{\framing}[1]{{\sffamily\color[HTML]{1B7A43} #1}}
    
    \newcommand{\variant}[2]{{\sffamily\color{variant#1}[alt #1] #2}}
    \newenvironment{variantblock}[1]{\par\sffamily\color{variant#1}[alt #1] }{\par}
    \newcommand{\addcite}[1]{\textsf{\color{red!80!black}[\#]}}
\else
    \renewcommand{\sout}[1]{}
    
    \newcommand{\framing}[1]{}
    
    \newcommand{\variant}[2]{}
    
    \newcommand{\addcite}[1]{}
\fi
\ifshowauthornotes
    \newcommand{\aaron}[1]{{\sffamily\color[HTML]{6E3CBC}(AS: #1)}}
    \newcommand{\zack}[1]{{\sffamily\color[HTML]{8378A2}(ZG: #1)}}
    \newcommand{\old}[1]{{\sffamily\color{gray}(old: #1)}}
\else
    \newcommand{\aaron}[1]{}
    \newcommand{\zack}[1]{}
    \newcommand{\old}[1]{}
\fi
\ifshowreviewernotes
    \newcommand{\sandy}[1]{{\sffamily\color[HTML]{D2691E}(SK: #1)}}
    \newcommand{\iz}[1]{{\sffamily\color[HTML]{8378A2}(IZ: #1)}}
    \newcommand{\nate}[1]{{\sffamily\color[HTML]{289FB8}(NB: #1)}}
    \newcommand{\dan}[1]{{\sffamily\color{magenta}(DP: #1)}}
    \newcommand{\jk}[1]{{\sffamily\color[HTML]{B8860B}(JK: #1)}}
    \newcommand{\jm}[1]{{\sffamily\color[HTML]{009c50}(JM: #1)}}
    \newcommand{\mk}[1]{{\sffamily\color[HTML]{1F6FB2}(MK: #1)}}
    \newcommand{\jh}[1]{{\sffamily\color[HTML]{00838F}(JH: #1)}}
\else
    \newcommand{\sandy}[1]{}
    \newcommand{\iz}[1]{}
    \newcommand{\nate}[1]{}
    \newcommand{\dan}[1]{}
    \newcommand{\jk}[1]{}
    \newcommand{\jm}[1]{}
    \newcommand{\mk}[1]{}
    \newcommand{\jh}[1]{}
\fi
\newcommand{\bea}{\begin{equation} \begin{aligned}}
\newcommand{\eea}{\end{aligned} \end{equation} }
\newif\ifptitle
\newif\ifpnumber
\newcounter{para}

\makeatletter
\def\maketitle{
\@author@finish
\title@column\titleblock@produce
\suppressfloats[t]}
\makeatother

\usepackage{titletoc}
\usepackage{tocloft}
\newcommand{\beginsupplement}{
      \setcounter{table}{0}
      \renewcommand{\thetable}{S\arabic{table}}
      \renewcommand{\theHtable}{S\arabic{table}}
      \setcounter{figure}{0}
      \renewcommand{\thefigure}{S\arabic{figure}}
      \renewcommand{\theHfigure}{S\arabic{figure}}
      \setcounter{section}{0}
      \renewcommand{\thesection}{S\arabic{section}}
      \renewcommand{\theHsection}{S\arabic{section}}
}

\newcommand{\makesititle}{Supplementary Information: \@title}

\newcommand{\figfont}{\fontfamily{DejaVuSans-TLF}\selectfont}
\newcommand{\panellabel}[1]{{\figfont\fontsize{8}{10}\selectfont\bfseries #1}}

\begin{document}
\renewcommand{\appendixname}{}
\title{HERA: Assessing an Autonomous Agent for Quantum Device Measurement}

\def\StanfordPhys{Department of Physics, Stanford University, Stanford, CA 94305}
\def\SIMES{Stanford Institute for Materials and Energy Sciences, SLAC National Accelerator Laboratory, Menlo Park, CA 94025}

\author{Zachary~W.~Gomez}
\email{zackg@stanford.edu}
\affiliation{\SIMES}
\affiliation{\StanfordPhys}

\author{Nathan~Bittner}
\affiliation{\StanfordPhys}

\author{Jordan~T.~McCourt}
\affiliation{\SIMES}
\affiliation{\StanfordPhys}

\author{Qingrui~Cao}
\affiliation{\SIMES}
\affiliation{\StanfordPhys}

\author{Sandesh~S.~Kalantre}
\affiliation{\StanfordPhys}

\author{Benjamin~H.~Alexander}
\affiliation{\StanfordPhys}

\author{Zhiren~Zheng}
\affiliation{\StanfordPhys}

\author{Kenji~Watanabe}
\affiliation{Research Center for Electronic and Optical Materials, National Institute for Materials Science, 1-1 Namiki, Tsukuba 305-0044, Japan}

\author{Takashi~Taniguchi}
\affiliation{Research Center for Materials Nanoarchitectonics, National Institute for Materials Science,  1-1 Namiki, Tsukuba 305-0044, Japan}

\author{Aaron~Sharpe}
\email{aaron.sharpe@stanford.edu}
\affiliation{\SIMES}
\affiliation{\StanfordPhys}

\date{\today}

\begin{abstract}

Large language model agents have begun to carry out laboratory measurements, but their scientific conclusions are useful only if they hold up under scrutiny.
Demonstrating this requires controlled studies because models are stochastic and the context a model is given steers its conclusions.
Here we introduce \textsc{HERA}, an autonomous agent designed to carry out cryogenic transport measurements of quantum devices.
\textsc{HERA} draws on a knowledge base curated by lab members that holds measurement protocols, software tools, and literature references.
We assessed \textsc{HERA} with a blinded trial against an expert on the stacking order of two multilayer graphene devices, an ablation of its knowledge base, and hundreds of offline replays in which models designed the trial's central measurement without running it.
\textsc{HERA}'s conclusions matched the expert's sealed conclusions on both tested devices. 
The ablation and the replay studies measured the impact of the knowledge base across different models, showing that with it every model designed better measurements and reasoned more soundly.
To test how readily \textsc{HERA} extends to a new problem, we tasked it with finding and characterizing superconductivity in a different material system.
This required a few hours of expert time to develop a new experimental protocol, after which \textsc{HERA} did both autonomously.
These results show that, with a knowledge base designed for the task, language model agents are capable of sound experimental science.

\end{abstract}

\maketitle

\section{Introduction}\label{sec:main:intro}

Large language models (LLMs) are being applied to experimental work across the sciences~\cite{Boiko2023, Shi2026qumus, Cao2024kagents, Qiu2026aims, Bauer2026fermilink}.
Here, we seek to automate experiments on quantum devices assembled from atomically thin materials, where a measurement records a device's response as a function of many parameters, including carrier density, displacement field, magnetic field, and temperature.
Together these axes span a parameter space too high-dimensional to search exhaustively.
Rather, testing a hypothesis about a device takes a campaign of targeted measurements.
Each interpretation builds on those before it, so the quality of every interpretation compounds into the conclusion.
Many experiments across the sciences share this structure.

For an LLM to run an experiment autonomously, it must keep the device and measurement rig safe, take the measurements needed to answer a question, and draw conclusions supported by the data.
An LLM's general knowledge is not enough to run a measurement campaign.
It needs the specifics of the lab's own hardware, procedures, and materials, as well as a connection to the measurement rig.
Here we introduce \textsc{HERA} (Hierarchical Experimentation and Reasoning Architecture).
Beyond augmenting the Claude Code harness with tools to control the lab instruments, it supplies the agent with a knowledge base curated by lab members, hereafter referred to as the corpus.
The corpus is a version-controlled collection of measurement protocols, software tools, and literature references built from lab notebooks and published literature.

Understanding the context provided to a model is critical because it steers the model's conclusions.
Changes to the context or tools given can alter its behavior in unintended ways that are often difficult to quantify.
A change intended to help may merely clutter the context.
Even when provided the same context, LLMs are stochastic, and a model run twice may return different outputs.
Assessing the performance of an LLM based system therefore requires careful controls.

To assess \textsc{HERA}, we ran four tests.
First, in a blinded prospective trial, \textsc{HERA} was asked to determine the stacking order of two multilayer graphene devices using cryogenic transport measurement.
For both devices, \textsc{HERA}'s conclusion matched an expert's sealed conclusion.
Second, we tested whether that agreement depends on what the corpus provides and on the model used.
In an ablation study, two different models each ran two campaigns, once with the full corpus and once with an ablated corpus that kept only its software tools (a ``corpus~$\times$~model'' matrix).
With the full corpus, both models agreed with the expert's stacking order, one at lower confidence.
With the ablated corpus, one model disagreed with the expert, and the other agreed but relied on claims its measurements do not support.
Third, we ran 300 ``offline replays'' in which a model was asked to choose the ranges and resolution of the trial's central measurement.
Six models designed the measurement repeatedly with either the full or the ablated corpus.
We scored each designed measurement from 0 to 1 by how well it could resolve the signatures that distinguish the candidate stacking orders.
Measurements designed with the full corpus scored higher on average than those designed with the ablated corpus, while also requiring less acquisition time.
Fourth, in a test of domain transfer, we examined how readily \textsc{HERA} extends to a new experimental question in a new setup.
With a new measurement protocol added to the corpus, it searched for superconductivity in a WSe$_2$-proximitized bilayer graphene device, using a second rig.
After two corrections to that protocol, it autonomously found and characterized superconductivity.
Taken together, these tests establish that \textsc{HERA} can conduct a measurement campaign through to conclusions that hold up under an expert's scrutiny.

\section{HERA}\label{sec:main:tactical}

\begin{figure*}[t]
    \centering
    \begin{tikzpicture}[xscale=1.0, yscale=0.85,   %
  font=\figfont\scriptsize,
  >={Triangle[length=3.4pt, width=3.4pt]},   %
  box/.style={draw, rounded corners=2pt, align=center, fill=white,
              inner sep=4pt, minimum height=2.2em},
  agentbox/.style={box, very thick, fill=heraAgent, minimum height=5.4em,
                   text width=10.6em},
  iobox/.style={box, text width=5.6em},
  hwbox/.style={box, fill=black!6, text width=4.8em},
  loopbox/.style={box, fill=heraLoop!12, text=heraLoop!60!black},
  group/.style={draw=heraGroup, dashed, thick, rounded corners=4pt, inner sep=6pt},
  grouplabel/.style={font=\figfont\scriptsize, text=heraGroup},
  sub/.style={font=\figfont\scriptsize, text=black!55},
  wire/.style={thick, ->},
  loopwire/.style={thick, ->, heraLoop, densely dashed},
  lens/.style={heraLoop, densely dashed, line width=0.5pt},
]

\node at (-7.4, 3.1) {\panellabel{a}};

\node[agentbox] (agent) at (0,0) {};
\node[anchor=north, font=\figfont\footnotesize\bfseries] at ([yshift=-5pt]agent.north) {HERA};

\node[iobox] (brief)   at (-4.93, 1.2)  {Campaign brief};
\node[iobox] (devcard) at (-4.93, 0)    {Device card};
\node[iobox] (fricard) at (-4.93,-1.2)  {Rig card};
\draw[wire, rounded corners=4pt] (brief.east) -| ([xshift=-0.35cm, yshift=0.3cm]agent.west) -- ([yshift=0.3cm]agent.west);
\draw[wire] (devcard.east) -- (agent.west);
\draw[wire, rounded corners=4pt] (fricard.east) -| ([xshift=-0.35cm, yshift=-0.3cm]agent.west) -- ([yshift=-0.3cm]agent.west);
\node[group, inner xsep=12pt, fit=(brief)(devcard)(fricard),
      label={[grouplabel]above:Session inputs}] (inputs) {};
\path let \p1 = (devcard.south), \p2 = (fricard.north) in
  node[box, text width=5.6em, inner sep=3pt, minimum height=1.8em, anchor=south]
  (prompt) at ([yshift=\y1-\y2]agent.south) {Agent prompt};

\node[iobox] (mnb)  at (4.93, 0.45)  {Measurement\\notebook};
\node[iobox] (raw)  at (4.93,-0.75)  {Raw data};
\node[iobox] (anb)  at (4.93,-1.95)  {Analysis\\notebook};
\node[iobox] (log)  at (4.93,-3.15)  {Session notes};
\node[iobox] (chat) at (4.93,-4.35)  {Chat\\transcript};
\draw[wire, <->] (agent.east |- mnb) -- (mnb.west);
\draw[wire] (mnb.south) -- (raw.north);
\draw[wire] (raw.south) -- (anb.north);
\coordinate (ftop) at ([yshift=0.35cm]mnb.north);
\node[group, inner xsep=12pt, fit=(mnb)(raw)(anb)(log)(chat)(ftop),   %
      label={[grouplabel, anchor=south east]north east:Session outputs}] (session) {};
\coordinate (r1) at ([xshift=-0.3cm]session.west);
\coordinate (r2) at ([xshift=-0.55cm]session.west);
\coordinate (r3) at ([xshift=-0.8cm]session.west);
\draw[wire, <->, rounded corners=4pt] ([yshift=0.15cm]agent.east) -| (r1 |- anb.west) -- (anb.west);
\draw[wire, rounded corners=4pt] ([yshift=-0.15cm]agent.east) -| (r2 |- log.west) -- (log.west);
\draw[wire, rounded corners=4pt] ([yshift=-0.45cm]agent.east) -| (r3 |- chat.west) -- (chat.west);

\path let \p1 = (mnb.south), \p2 = (raw.north) in
  node[hwbox, minimum height=0.9cm, text width=5.2em, anchor=south] (rig) at ([yshift=\y1-\y2]agent.north) {Measurement rig};
\draw[wire, <->, rounded corners=4pt] ([xshift=0.25cm]mnb.north west) |- (rig.east);

\tikzset{kbox/.style={box, font=\figfont\scriptsize, inner sep=3pt},
         toolbox/.style={box, font=\figfont\scriptsize, inner sep=3pt, fill=black!6},
         webwire/.style={black!40, line width=0.5pt}}
\begin{scope}[xshift=-1.55cm, yshift=-0.15cm]
\node[kbox] (rec1) at ( 0.1,-2.9)  {Recipe};
\node[kbox] (ref2) at ( 2.0,-2.9)  {Reference};
\node[kbox] (rec2) at (-1.0,-4.05) {Recipe};
\node[toolbox] (tool2) at ( 1.0,-4.05) {Tool};
\node[kbox] (rec3) at ( 2.8,-4.05) {Recipe};
\end{scope}
\draw[webwire] (rec1) -- (ref2);
\draw[webwire] (rec2) -- (tool2);
\draw[webwire] (tool2) -- (rec3);
\draw[webwire] (rec1) -- (rec2);
\draw[webwire] (rec1) -- (tool2);
\draw[webwire] (ref2) -- (rec3);
\draw[webwire] (ref2) -- (tool2);
\tikzset{more/.style={black!65, line width=0.8pt, dotted}}
\draw[more] (rec2.west) -- ++(-0.25,0);
\draw[more] (rec1.west) -- ++(-0.6,0);
\draw[more] (ref2.east) -- ++(0.25,0);
\draw[more] (rec3.east) -- ++(0.25,0);
\node[draw, dashed, rounded corners=3pt, fit=(rec1)(ref2)(rec2)(tool2)(rec3),
      inner sep=6pt, label={[font=\figfont\scriptsize, text=black!55, anchor=south west]north west:corpus (knowledge base)}]
      (knowledge) {};
\draw[wire] ([xshift=0.35cm]ref2 |- knowledge.north) -- ([xshift=0.35cm]ref2 |- agent.south);

\node[hwbox, text width=3.8em] (hist) at (-5.1,0 |- knowledge.west) {Historical\\data};
\draw[wire, black!55] (hist.east) -- (knowledge.west);
\node[sub, align=center, anchor=north] at (hist.south) {one-time\\ingestion};

\node[loopbox, text width=6.4em, anchor=north] (selfimp) at ([yshift=-0.55cm]knowledge.south) {Self-improvement\\loop};

\path let \p1 = (inputs.west), \p2 = (-7.25,0) in coordinate (busx) at ($(session.east)+(\x1-\x2,0)$);
\draw[loopwire, rounded corners=10pt]
  (session.east |- log) -- (busx |- log) -- (busx |- selfimp) -- (selfimp.east);

\draw[loopwire] (selfimp.north) -- (knowledge.south);  %
\coordinate (looptop) at ([yshift=0.25cm]rig.north);
\draw[loopwire, rounded corners=10pt]
  (selfimp.west) -- (-7.25,0 |- selfimp) -- (-7.25,0 |- looptop) -- (-1.35,0 |- looptop) -- (-1.35,0 |- agent.north);

\begin{scope}[yshift=-9.1cm]
\node at (-7.4, 0.6) {\panellabel{b}};

\node[box, text width=5.2em, draw=heraGroup, thick] (run) at (-4.2, 0) {Session};
\node[box, text width=5.2em] (archive) at (-1.4, 0) {Session\\archive};
\node[box, text width=5.2em] (lessons) at ( 1.4, 0) {Findings};
\node[box, dashed, text width=5.2em] (updates) at ( 4.2, 0) {Corpus\\revisions};

\node[box, text width=5.2em] (fixlogback) at (1.52,-2.02) {\phantom{Fix log}};
\node[box, text width=5.2em] at (1.46,-1.96) {\phantom{Fix log}};
\node[box, text width=5.2em] (fixlog) at (1.4,-1.9) {Fix log};
\node[sub, anchor=north] (accum) at (1.4,-2.55) {accumulates across sessions};

\draw[wire] (run) -- (archive);
\draw[wire] (archive) -- (lessons);
\draw[wire] (lessons) -- (updates);

\draw[wire] (lessons.250) -- node[left, sub] {deferred} (fixlog.110);
\draw[wire] (fixlog.70) -- node[right, sub] {recurrence} (lessons.290);
\draw[wire, rounded corners=6pt] (updates.south) |-
  node[pos=0.75, below, sub] {recorded} (fixlogback.east |- fixlog.east);

\draw[loopwire] (updates.north) to[out=120, in=60, looseness=0.35]
  node[midway, above, sub, text=heraLoop] (smarter) {next session starts smarter} (run.north);

\node[draw=heraLoop, densely dashed, thick, rounded corners=4pt, inner sep=8pt,
      fit=(run)(updates)(accum)(smarter)] (loopframe) {};
\end{scope}
\draw[lens] (selfimp.south west) -- (loopframe.north west);
\draw[lens] (selfimp.south east) -- (loopframe.north east);

\end{tikzpicture}
    \caption{\textbf{HERA.}
    (\textbf{a}) HERA is an LLM agent designed to conduct autonomous cryogenic transport measurements on quantum devices.
    The agent prompt holds the standing instructions HERA follows in every measurement session.
    A session starts from its inputs: a campaign brief, a device card, and a rig card.
    The brief states the campaign's goals.
    The cards are machine-readable datasheets for the device being measured and the rig measuring it.
    HERA draws on a corpus of recipes, software tools, and literature references.
    A recipe outlines an experimental protocol and points to the examples, tools, and related recipes.
    The corpus was seeded from the lab's historical data.
    HERA writes code into two notebooks, one for measurement and one for analysis.
    Code in the measurement notebook drives the rig and records the raw data.
    HERA analyzes the raw data and reports its results in the analysis notebook.
    HERA writes problems it encounters and opportunities for improvement to its session notes.
    The chat transcript records every action HERA takes and every message it exchanges with the operator.
    The notebooks, raw data, session notes, and chat transcript are the session's outputs.
    Together these form the session archive, which feeds the self-improvement loop.
    (\textbf{b})~After each session, the operator launches a review agent that reads the archive for findings and checks it against the fix log.
    This check validates previous fixes and escalates small issues that recur.
    Each finding becomes a proposed revision to the corpus or the agent prompt, or is deferred to the fix log.
    The operator accepts or rejects each proposed revision, and the fix log records each revision.
    }
    \label{fig:main:1}
\end{figure*}
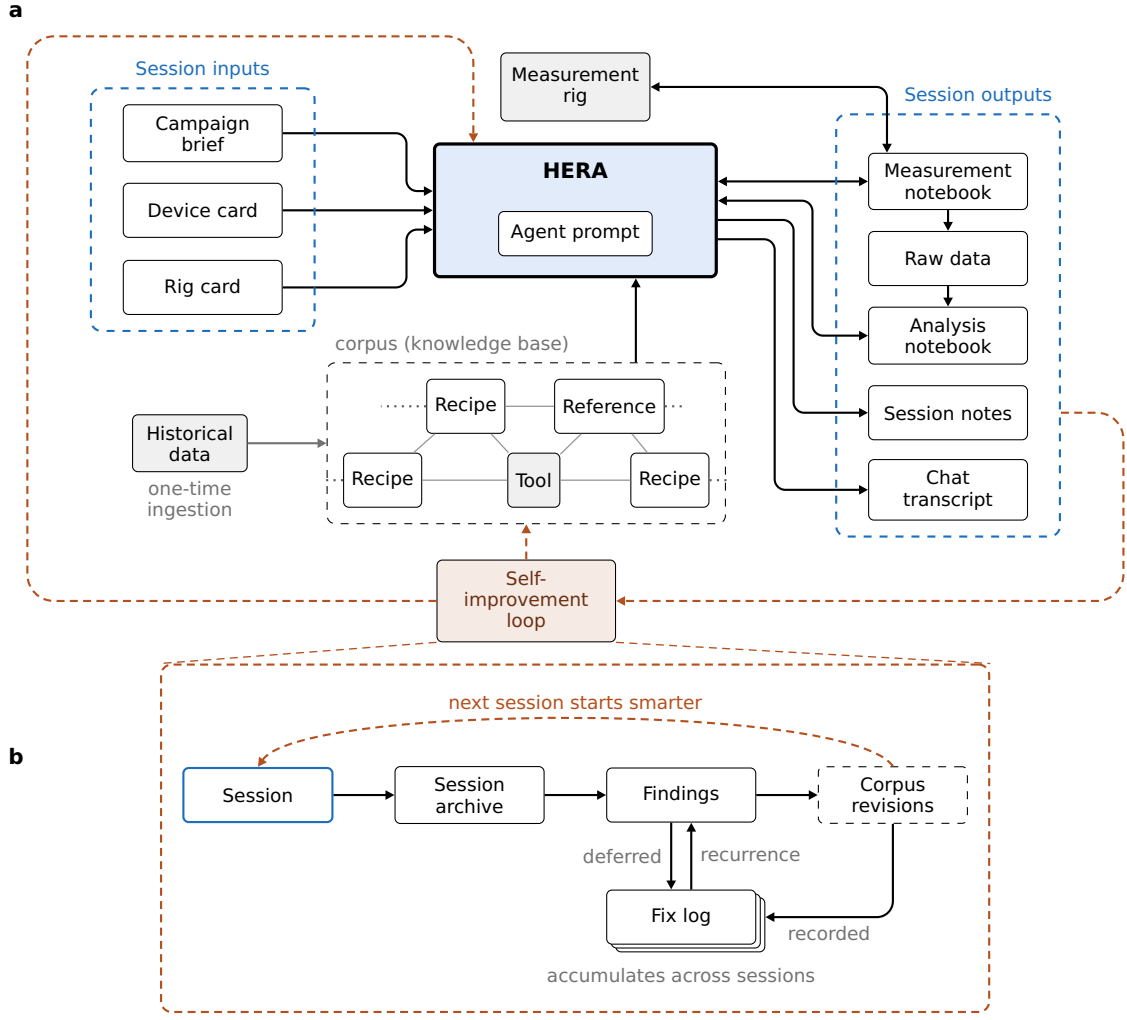

At the center of \textsc{HERA} is its curated corpus of recipes, software tools, and literature references (Fig.~\ref{fig:main:1}a).
By design, the corpus provides resources rather than instructions to execute. 
They are synthesized in collaboration with an LLM, drawing from the literature, the lab's accumulated knowledge, and the self-improvement loop described below.
A recipe outlines an experimental protocol, when to use it, what it measures, and heuristics for carrying it out.
Each recipe points to examples, tools, and related recipes, so the agent can move from one recipe to the next as the data suggest.
Software tools are the clearest case of a resource rather than an instruction.
They package the analyses that recipes call for.
For example, a single tool filters a signal and computes its Fourier transform.
The defined input and output of a tool forces the agent to acquire and provide the necessary inputs and consume the validated and curated outputs, creating a ``pit of success'' where the easiest path is the correct path.
The self-improvement loop adds more where it finds a repeated or error-prone pattern.

A measurement session begins with a campaign brief from the human operator that defines the goals, boundaries, and potential starting points. 
The brief points to a device card and a rig card, machine-readable datasheets for the device being measured and the rig measuring it  (Fig.~\ref{fig:main:1}a).
In the measurements we discuss below, the device card gives the device's identifier, composition, geometry, wiring, and safety limits. 
The rig card gives the rig's instrument loadout, operating limits, and quirks.
Everything that follows --- which recipes to consult, what data to acquire --- is the agent's choice.
Regardless of the path, the agent must answer the questions posed in the brief at one of a fixed set of confidence levels.
A guide in the corpus defines the confidence levels, and each recipe sets what evidence constitutes each level (Methods).

The measurement agent conducts the campaign by writing code and comments into two live Jupyter notebooks, one for measurement and one for analysis.
In the measurement notebook, the agent designs each measurement and writes the code that drives the rig and records the raw data.
Alongside the code, it records the goal of each measurement and its reasoning.
Whether a measurement runs as soon as the agent writes its code, or waits for the operator to run it, is set by the level of autonomy defined for each session (Methods).
In the analysis notebook, the agent plots the data, performs any relevant analysis, and states its conclusions.
The notebooks and raw data are the scientific output of the session.
These Jupyter notebooks follow how the lab already executes and documents its measurements, so an operator can read a session as they would their own and can easily step in to assist or oversee the agent.
As the agent works, it keeps session notes on the problems it encounters and the opportunities for improvement it identifies for use in the self-improvement loop.
The chat transcript, saved automatically by the agent harness, holds all remaining details: every action the agent took and every message it exchanged with the operator.
A finished session is assembled into an archive containing the input brief, cards, notebooks, raw data, session notes, and transcript.
The archives of all sessions in this work are publicly available (Data availability).

After each session, the operator launches a review agent that reads the archive for findings and proposes revisions (Fig.~\ref{fig:main:1}b).
The operator accepts or rejects each proposed revision, and an accepted finding becomes a revision to the corpus or an entry in a fix log.
Because the log accumulates across sessions, revisions can come from a pattern that recurs across many sessions as well as single findings.
The HERA system --- corpus and agent prompt --- is kept under version control, so every revision is recorded and any past state can be checked out and run again.

\begin{figure*}[t]
    \centering
    \input{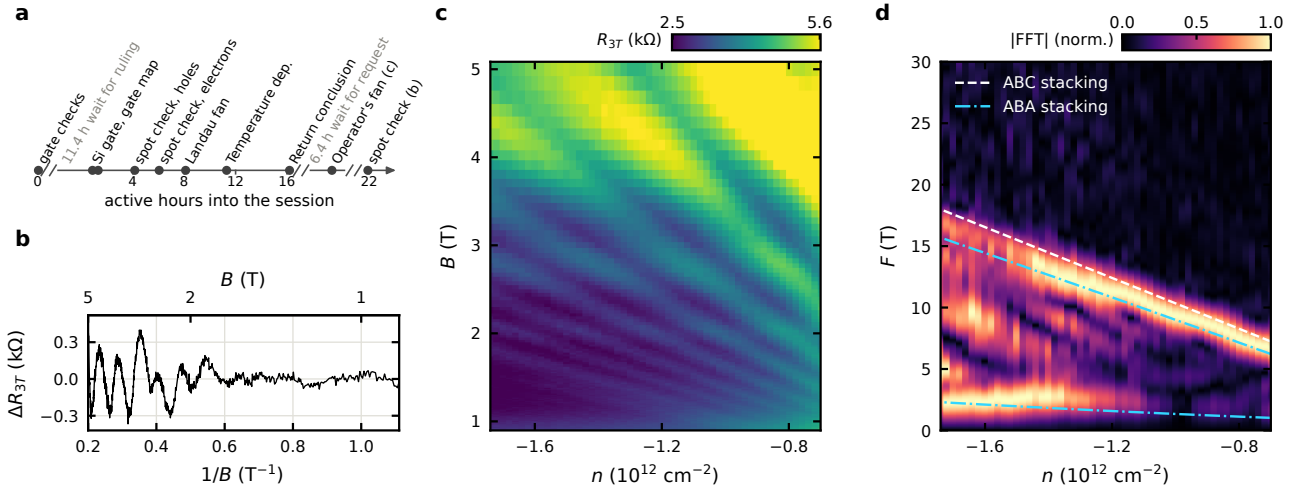}  
    \caption{
    \textbf{\textsc{HERA}'s measurement sequence for stacking order}
    (\textbf{a}) Timeline of HERA's session on the trilayer device, from the brief to the spot check in (\textbf{b}), in active hours.
    The first break in the axis excludes an 11.4\,h wait for the operator's ruling on a safety concern HERA raised.
    The second break excludes the 6.4\,h between HERA's conclusion and the operator's request for the fan in (\textbf{c}).
    (\textbf{b}) Background-subtracted three-terminal resistance $\Delta R_{3T}$ as a function of inverse magnetic field $1/B$ at carrier density $n = -1.74\times10^{12}$~cm$^{-2}$ and displacement field $D = 0$.
    In this spot check, $\Delta R_{3T}$ exhibits Shubnikov--de Haas (SdH) oscillations, which are used to discriminate between stacking orders (Sec.~\ref{sec:si:stacking}).
    (\textbf{c}) Three-terminal resistance $R_{3T}$ as a function of $n$ and $B$ at $D = 0$.
    HERA measured this higher-resolution follow-up Landau fan at the operator's request after its conclusion.
    Fig.~\ref{fig:si:trial_fans} shows the fan HERA concluded from and the expert's fan.
    (\textbf{d}) Fourier amplitude $|\mathrm{FFT}|$ of each constant-density linecut of (\textbf{c}), each normalized to its maximum, as a function of $n$ and SdH frequency $F$.
    The lines indicate the frequencies expected for ABC and ABA stacking at the density calibration HERA set in the session.
    }
    \label{fig:main:2}
\end{figure*}

\section{A blinded prospective trial}\label{sec:main:trial}

We first test whether \textsc{HERA} can reach the same conclusion as an expert~\footnote{The expert is A.S.} on the same question about a device.
To create a fair and unbiased test, the expert's conclusion was sealed before each session and withheld from both \textsc{HERA} and its operator~\footnote{The operator is Z.W.G.}.
A protocol written in advance for the trial fixed what was withheld, the form \textsc{HERA}'s conclusion had to take, and how it would be compared with the expert's (Methods).
The protocol also required every session to be published, so no favorable session could be chosen after the fact. 
As we show below, \textsc{HERA}'s conclusion agreed with the expert's for both devices in the trial.

Both \textsc{HERA} and the expert were asked to determine the stacking order of the two multilayer graphene devices in the trial.
Multilayer graphene is a stack of a few graphene sheets exfoliated from graphite (Methods).
It can have one of several stacking orders, which differ in how each sheet is laterally offset from the one below~\cite{ViznerStern2025polytypes, Lui2011imaging}.
Bernal stacking (AB) is more stable than rhombohedral stacking (ABC)~\cite{Bernal1924, Lipson1942}.
Rhombohedral stacking is nonetheless a focus of current research, because its low-energy band edges are flat~\cite{Min2008chiral, Koshino2009gate, Zhang2010abc}, making it a platform for superconductivity and other strongly correlated phases~\cite{Chen2019mott, Chen2019superconductivity, Chen2020chern, Zhou2021half, Zhou2021superconductivity, Lu2024fractional, Han2025chiral, Kalantre2026fermiology}.
Strain during device fabrication can convert rhombohedral stacking to Bernal stacking~\cite{Geisenhof2019soliton, Yang2019stacking}, so the stacking order of a finished device has to be measured, and charge transport is a standard way to do so~\cite{Bao2011stacking}.
How the stacking order is determined from transport is described in more detail in Sec.~\ref{sec:si:stacking}.

In this blinded prospective trial, HERA was given two multilayer graphene devices to measure: one trilayer (device 1) and one tetralayer (device 2).
Both devices were assembled from flakes in which a rhombohedral domain had been isolated before stacking (Methods).
Two gates, one above and one below the stack, tune the carrier density $n$ and the displacement field $D$ in each device (Methods).

For both \textsc{HERA} and the expert, each conclusion on a device's stacking order had to be stated at one of a fixed set of confidence levels (Methods).
After the initial development on a different device utilizing the self-improvement loop (Sec.~\ref{sec:si:revisions}), the HERA system was frozen for this and later trials.
Before the trial, the expert measured each device and independently concluded its stacking order.
This conclusion was sealed and withheld from both \textsc{HERA} and its operator (Methods). 

HERA received each device already cold in the cryostat and ready to measure.
Its only inputs were the brief, the device card, and the rig card.
The brief asked which of the possible stacking orders each device has, ABA or ABC for the trilayer device and ABAB, ABCA, or ABCB for the tetralayer device.
From there HERA proceeded autonomously, measuring on its own until it was ready to state its conclusion.
Its sequence of measurements on the trilayer device (Fig.~\ref{fig:main:2}a) was typical of its stacking-order sessions with the corpus both here and in the next section (Sec.~\ref{sec:main:matrix}).

The sequence centered on a Landau fan (Fig.~\ref{fig:main:2}c), the resistance measured as a function of $n$ and magnetic field $B$.
Each constant-density linecut of the fan can show Shubnikov--de Haas (SdH) oscillations, oscillations of the resistance periodic in $1/B$ (Fig.~\ref{fig:main:2}b).
Each stacking order exhibits a distinct set of SdH frequencies, so the measured frequencies can distinguish between the orders (Sec.~\ref{sec:si:stacking}).

A Landau fan is a time-consuming measurement, with a typical measurement lasting many hours, so the recipe for determining stacking order suggests a sequence of checks beforehand that inform the design of the fan (Fig.~\ref{fig:main:2}a).
Initial gate checks confirm that the device is healthy and calibrate parameters needed for later measurements and analyses.
Next, SdH spot checks measure the resistance as a function of $B$ at fixed $n$ and $D$ (Fig.~\ref{fig:main:2}b).
They find the onset field of SdH oscillations and how cleanly the oscillations are resolved, which guide the field and density ranges of the fan.
In the trilayer session, HERA took a fan on the hole side, where the spot checks showed cleaner oscillations with a lower onset field (Sec.~\ref{sec:si:stacking}).
ABC stacking is expected to exhibit a single SdH frequency, whereas ABA is expected to exhibit a pair of frequencies (Sec.~\ref{sec:si:stacking}).
The fan resolved a single frequency consistent with the higher of the two ABA frequencies under the assumed gate calibration (Fig.~\ref{fig:si:trial_fans}).
At any density in the fan, the lower ABA frequency would complete at most one oscillation across the fan's field range, too few to resolve.
According to the rules of evidence in the corpus, HERA therefore counted it as unresolved, not absent.

\begin{figure*}[t]
    \centering
    \input{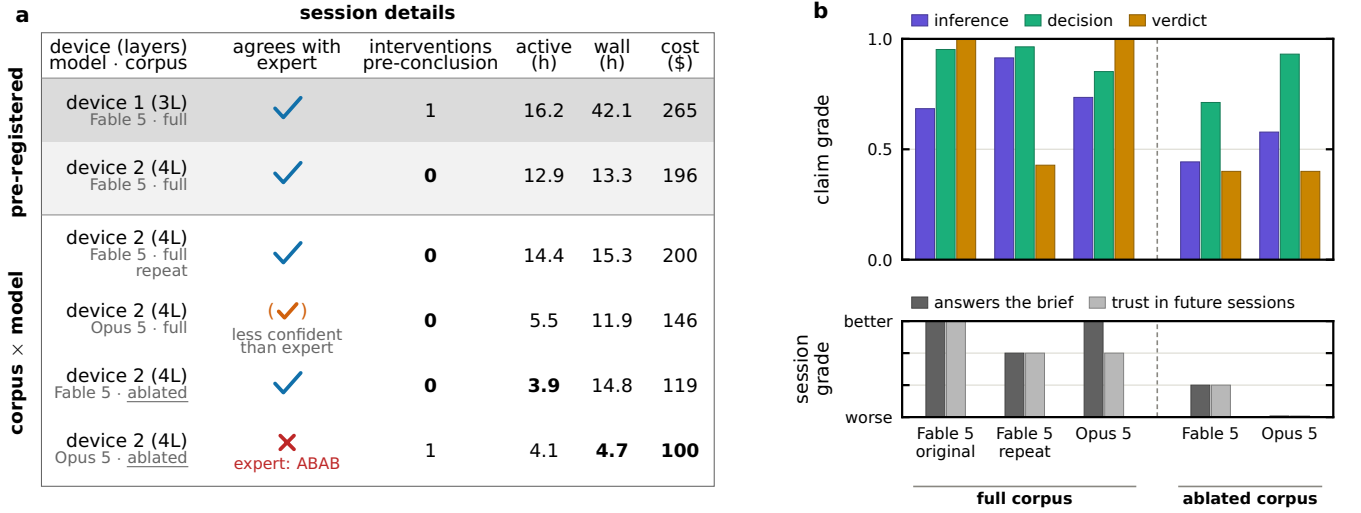}
    \caption{\textbf{Live sessions with the full and ablated corpus.}
    (\textbf{a}) The six live sessions asked to determine the stacking order of a multilayer graphene device.
    The two shaded sessions are the blinded prospective trial (Sec.~\ref{sec:main:trial}).
    The four unshaded sessions are the ablation study (Sec.~\ref{sec:main:matrix}).
    The ablated corpus kept only the software tools of the full corpus.
    Columns show agreement with the expert's sealed conclusion, operator interventions before the conclusion, active and wall-clock time, and cost at API list prices (Methods).
    Bold indicates the best value in each column.
    The interventions for all sessions are listed in Sec.~\ref{sec:si:interventions}.
    (\textbf{b}) Expert grades of the five tetralayer sessions.
    Top, the mean grade of each session's inference, decision, and verdict claims (Methods).
    Bottom, how well each session answers the brief and how much the expert would trust the same agent in future sessions.
    Both grades use the rubric's four options, from worse to better (Methods).
    Table~\ref{tab:si:trial_table} shows the claim score and both whole-session grades of every session.
    Fig.~\ref{fig:si:chain} shows the expert's ruling on each step of each session.
    For Opus 5 with the full corpus, the grades in (\textbf{b}) cover the whole session, including a finer Landau fan the operator requested after its first conclusion.
    Panel (\textbf{a}) shows its first conclusion, ABAB at lower confidence than the expert's, and its final conclusion after the finer fan was ABAB at the expert's confidence.
    }
    \label{fig:main:matrix}
\end{figure*}

With the lower ABA frequency unresolved, HERA turned to a second discriminator suggested by the recipe, the mass of the carriers, which it extracted from how the SdH oscillations weaken as a function of increasing temperature (Sec.~\ref{sec:si:stacking}).
The extracted mass was close to the value expected for ABA and far from that expected for ABC.
With the mass and the fan's frequency both pointing to ABA, HERA concluded ABA.
The operator, blind to the sealed conclusion, asked for a finer fan extending to lower field in an attempt to resolve the lower frequency expected for ABA (Fig.~\ref{fig:main:2}c,d).
The finer fan resolved the lower ABA frequency but left HERA's conclusion unchanged in stacking order and confidence level (Fig.~\ref{fig:main:2}d).

The expert's sealed conclusion was then opened, and in both sessions of the trial, HERA's conclusion matched the expert's in both stacking order and confidence level.
\textsc{HERA} drew both conclusions autonomously, without steering from the operator.
The tetralayer session ran without a single intervention from the operator.
The trilayer session had one intervention before its conclusion, about a safety concern HERA raised over gate leakage (Fig.~\ref{fig:main:2}a), and one after, the request for the finer fan.
The interventions for all sessions are listed in Sec.~\ref{sec:si:interventions}.

\section{Corpus $\mathbf{\times}$ model}\label{sec:main:matrix}

We now test whether the agreement in the trial depends on what the corpus provides and on the model used.
Two models, Fable 5 and Opus 5, each measured the tetralayer device twice, once with the full corpus and once with an ablated corpus that kept only its software tools.
Both conditions had the same goal, device card, and rig card; the ablated brief named no recipes and asked the agent to write its own measurement plan.
Because a conclusion alone says little about how a session reached it, the expert also graded each session on its reasoning.
As we show below, with the full corpus, both models agreed with the expert.
With the ablated corpus, one model disagreed with the expert's stacking order and the other agreed.
In both conditions the models planned reasonable measurements, but with the ablated corpus both took insufficient data to substantiate their claims.

The session that disagreed with the expert was run by Opus 5 with the ablated corpus, which concluded that the stacking order is ABCA.
Its conclusion relied on the false premise that Bernal tetralayer stays metallic at densities near the charge neutrality point under zero displacement field~\cite{Koshino2011landau, Shi2018lifshitz}.
The two sessions with the ablated corpus were the shortest at 3.9 and 4.1 active hours.
The Fable 5 sessions with the full corpus took 13 and 14 hours, where most of this extra time came from SdH spot checks and the temperature dependence of the oscillations.
Opus 5 with the full corpus took only 5.5 hours, because it stopped before completing the full set of measurements the recipe suggests and gave no reason for stopping.
It still concluded ABAB but at a lower confidence level than the expert's.

To assess how the sessions reached their conclusions, the expert graded each session's reasoning claim by claim and each session as a whole (Fig.~\ref{fig:main:matrix}b).
For the claim-by-claim grading, an LLM extracted the substantive claims each session made in its notebooks and chat transcript (Sec.~\ref{sec:main:tactical}), 151 claims in all (Methods).
Each claim was categorized as an inference from the data, a decision about what to measure, or a final verdict.
The claims were pooled and shuffled so the expert could not tell which session a claim came from.
Each claim was broken into the separate verifiable statements it makes, 561 items in all, and the expert graded each item as correct, partly correct, or wrong.
We compare the grades of the five sessions on the tetralayer device.
The sessions with the ablated corpus have the lowest inference score of the five sessions, so they misread their data more often than the sessions with the full corpus (Fig.~\ref{fig:main:matrix}b).
Each verdict was graded on the conclusion it drew and on how well the data substantiated it.
With the full corpus, the expert graded every verdict and its evidence correct, except in the Fable 5 repeat, discussed below.
In both sessions with the ablated corpus, the verdict rested on four lines of evidence, two of which were graded wrong.
The decisions, about what to measure and how, were graded high in all five sessions, so both models chose what to measure soundly under both conditions.

To check the consistency of the claim-by-claim grading, a second expert~\footnote{The second expert is S.S.K.} scored a fixed subset of 56 claims, blind to session identity (Methods).
The two experts gave the same grade on 65\% of the 145 items both scored.
The quadratic-weighted Cohen's $\kappa$, a measure of agreement on a scale from 0 for chance to 1 for full agreement, was 0.38 (Methods).
Most of the disagreement came from how strictly each expert graded, with the second expert the more lenient, and about 80\% of the disagreements were single-step changes in the grade.
The largest share fell on Fable 5's session with the ablated corpus, whose claims the second expert graded higher.
The difference in strictness did not change the separation between the full and ablated corpus, because under the second expert's grades every session with the corpus still outscored both sessions with the ablated corpus (Sec.~\ref{sec:si:second_expert}).

To grade each session as a whole, the expert read both its measurement and analysis notebooks in full and gave the session two grades: one for whether the work answers the brief as it stands, after re-analysis, after re-measurement, or not at all, and one for how much supervision the same agent would need in a further session (Fig.~\ref{fig:main:matrix}b).
The whole-session grades clearly separate the two conditions.
Both sessions with the ablated corpus were graded as needing re-measurement or as not answering the brief.
Two of the three sessions with the corpus were graded as answering the brief as is, and the Fable 5 repeat as answering it after re-analysis.
The grades of Opus 5 with the corpus include a fan the operator requested after its first conclusion, because only its final verdict was graded, though both verdicts named ABAB (Sec.~\ref{sec:si:interventions}).
Under both conditions, the expert's most common criticism was that the sessions should have taken higher-resolution data.

\begin{figure*}[t]
    \centering
    \input{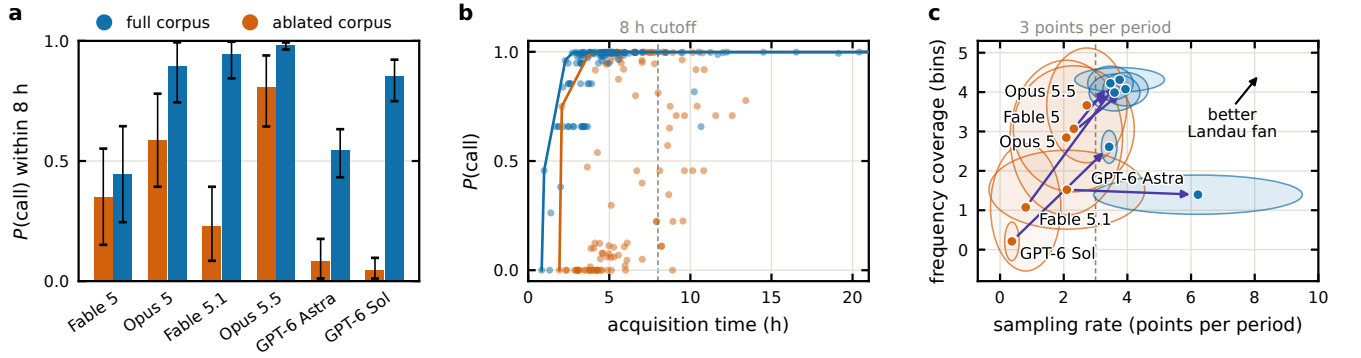}
    \caption{\textbf{Offline replay of Landau fan design.}
    (\textbf{a}) $P(\text{call})$ for six models, each designing up to twenty fans with the full corpus and up to twenty with the ablated corpus (Methods).
    $P(\text{call})$ estimates whether a fan would discriminate between multilayer graphene stacking orders.
    A fan planned to run longer than an eight-hour cutoff counts as zero.
    Bars show each model's mean with a 95\% bootstrap interval.
    (\textbf{b}) $P(\text{call})$ of all fans as a function of their planned acquisition time.
    Lines mark the Pareto frontiers of the fans with the full and the ablated corpus.
    A fan is on the frontier when no other fan in its group scores higher in less time.
    The dashed line indicates the eight-hour cutoff applied in (\textbf{a}).
    (\textbf{c}) Frequency coverage as a function of sampling rate for each model with the full and the ablated corpus.
    Sampling rate is the number of field points per period of the highest expected SdH frequency on the fan's highest-density linecut.
    Frequency coverage is the number of Fourier-transform frequency bins that the lowest expected SdH frequency covers across the linecuts that resolve it.
    Points give each model's mean and ellipses one standard deviation, and arrows run from the ablated corpus to the full corpus.
    The dashed line indicates the three points per period that the corpus's Landau fan recipe recommends.}
    \label{fig:main:benchmark}
\end{figure*}

Though an ABAB device is expected to show resistance peaks flanking charge neutrality~\cite{Shi2018lifshitz, Hirahara2018peaks}, no recipe of the corpus covers this feature.
Every agent in a tetralayer session initially misinterpreted these peaks as a signature of two regions of the device at different carrier densities rather than as a consequence of the band structure.
Opus 5 with the ablated corpus mistook one of the peaks for charge neutrality and read its rise under displacement field as indicating a gap that would rule out ABAB.
In the Fable 5 repeat, the two-region misinterpretation carried into the final verdict.
Its verdict on the stacking order agreed with the expert, but the expert graded the two-region claim wrong, which lowered the session's verdict grade (Fig.~\ref{fig:main:matrix}b).
With the full corpus, every session based its conclusion of the stacking order on the SdH frequencies the recipe suggests.
The corpus did not prevent the misreading of a feature it does not cover, but by suggesting which evidence to weight, it kept the peaks out of each session's stacking-order conclusion.

\section{Offline replay}\label{sec:main:replay}

We now extend the comparison of the full and ablated corpus to more models and more repetitions than live sessions allow by testing how each model designs a measurement offline.
Each live session takes hours of cryostat time, so the comparison of Sec.~\ref{sec:main:matrix} has only one or two sessions for each model and condition.
In an offline replay, a model is given the record of a live session up to the moment where that session previously designed its Landau fan and is asked to design the fan again.
The fan is never measured, so the replay can be run without access to the cryostat.
We replayed the design of the Landau fan because every stacking-order conclusion drawn with the full corpus drew on the SdH frequencies extracted from the fan.
We scored each design by how well it could distinguish the stacking orders.
As we show below, each of six models designed a better fan with the full corpus than with the ablated corpus, and the designs made with the full corpus required less acquisition time.

Each replay starts from the Fable 5 corpus repeat on the tetralayer device (Sec.~\ref{sec:main:matrix}), just after its two SdH spot checks (compare Fig.~\ref{fig:main:2}a), with a request to the model to design a fan.
The design sets the range of carrier density and magnetic field the fan covers, and how finely each range is sampled.
The replay ends when the model submits its fan design.
Six different models (Fable 5, Fable 5.1, Opus 5, Opus 5.5, GPT-6 Sol, and GPT-6 Astra) each designed up to twenty fans with the full corpus and up to twenty with the ablated corpus (Methods).
Of the 300 replays, 239 ended in a fan the grader could score.
The other 61 ended without a fan, almost always because the model designed another spot check instead (Methods).
All but two of the 61 were with the full corpus, whose recipe emphasizes spot checks before a fan is designed.

We scored each designed fan with a deterministic grader by how well it could resolve the SdH frequencies expected at the sampled densities for Bernal and rhombohedral stacking (Methods).
The grader was developed on fans from replays of different live sessions or models and frozen before any production fan was graded (Methods).
Its main metric, $P(\text{call})$, is the grader's score, from 0 to 1, for whether a fan as designed could discriminate between stacking orders.
We first compare the two conditions on the pre-registered statistic, the mean gain in $P(\text{call})$ across models with an eight-hour cutoff, counting any fan designed to run longer as zero.
The pooled gain does not depend on the choice of cutoff (Methods).
With this cutoff, every model's fans scored higher on average on $P(\text{call})$ with the full corpus than with the ablated corpus (Fig.~\ref{fig:main:benchmark}a).
The gain is statistically significant on a permutation test, $p = 0.0004$ after a Holm correction for the four metrics tested (Methods).
Opus 5.5's fans with the ablated corpus already scored well, so the corpus had relatively smaller impact.
The gains were largest for the two GPT models and Fable 5.1, whose fans with the ablated corpus scored lowest.

\begin{figure*}[t]
    \centering
    \input{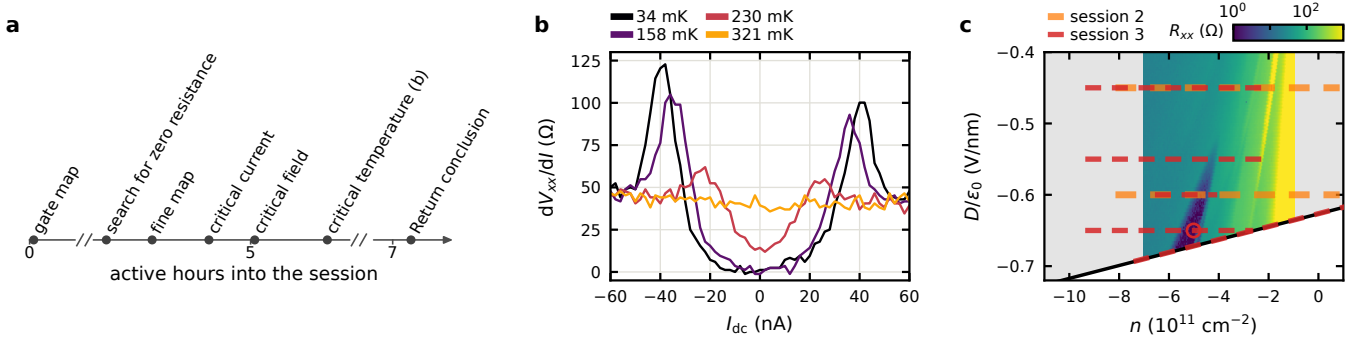}
    \caption{
    \textbf{HERA's search for superconductivity.}
    (\textbf{a}) Timeline of session 6 on the WSe$_2$-proximitized bilayer graphene device, from the grant of autonomy to HERA's conclusion, in active hours.
    (\textbf{b}) Differential resistance $\mathrm{d}V_{xx}/\mathrm{d}I$ as a function of dc current $I_\mathrm{dc}$ at four temperatures, acquired autonomously by HERA in the session in (\textbf{a}).
    (\textbf{c}) Longitudinal resistance $R_{xx}$ as a function of $n$ and $D/\epsilon_0$, measured in the additional session that HERA ran with the operator in the loop (Sec.~\ref{sec:si:zf1}).
    The diagonal black line denotes the boundary of the accessible gate range.
    Grey regions denote accessible regions of gate space outside the measured map.
    Dashed lines denote the cuts along which session~2 (orange) and session~3 (red) searched for superconductivity.
    The red circle denotes session 3's chosen operating point for subsequent measurements.
    Session~2's cuts reached only $D/\epsilon_0 = -0.60$~V/nm, just above where $R_{xx}$ begins to vanish, and it concluded that the device does not superconduct.
    Session~3, running with an additional coverage rule, added the sweep along the gate limit.
    }
    \label{fig:main:transfer}
\end{figure*}

We now remove the cutoff and compare the two conditions on generated fans' $P(\text{call})$ as a function of estimated acquisition time (Fig.~\ref{fig:main:benchmark}b).
Across all models, the full corpus raised the mean $P(\text{call})$ and lowered the median estimated acquisition time of the planned fan from 5.5 h to 4.4 h (Fig.~\ref{fig:main:benchmark}b).
The number of fans planned to run over eight hours fell from 29 to 15.
Pairing one fan from each condition for the same model, the full-corpus fan scored higher on $P(\text{call})$ and took less time in 45\% of pairs, and the ablated-corpus fan did so in 5\%.
The difference, 0.40, is the Pareto dominance of the full corpus (Methods).

Two further metrics each score a single aspect of the fan.
First, ``sampling rate'' is the number of field points per period of the highest expected SdH frequency, on the fan's highest-density linecut.
Second, ``frequency coverage'' is the number of frequency bins of the Fourier transform that the lowest expected SdH frequency covers across all linecuts that resolve it.
With the full corpus, the models designed better fans, and designed them more consistently (Fig.~\ref{fig:main:benchmark}c).
The mean sampling rate was higher for every model, and the mean frequency coverage for every model except GPT-6 Astra.
The one-standard-deviation ellipse of every model's designs in these two metrics was smaller.
The four Claude models converged on nearly the same design, while the two GPT models each landed elsewhere.
This may be a fingerprint of HERA having been developed with Claude models.

\section{Domain transfer}\label{sec:main:transfer}

Finally, we test how readily \textsc{HERA} extends to a new question.
We asked \textsc{HERA} to find and characterize any superconductivity in a different material system measured on a different rig.
The harness, the agents, and the existing corpus carried over unchanged.
The transfer required a device card, a rig card, and a recipe for investigating superconductivity.
As we show below, after two corrections to the new recipe, \textsc{HERA} autonomously found and characterized the superconductor.

The test bed for this transfer is Bernal bilayer graphene proximitized by \ce{WSe2}, a material system known to superconduct~\cite{Zhang2023enhanced}.
The expert had previously confirmed that the device measured here superconducts in a narrow sliver of the $n$--$D$ plane (Fig.~\ref{fig:si:zf1_characterization}c).

The expert composed the three documents required for the transfer in a few hours.
The recipe encodes the expert's own standard of evidence for superconductivity and a suggested order of measurements, grounded in Refs.~\cite{Zhou2021superconductivity, Zhang2023enhanced}.
The evidence sought is a longitudinal resistance dropping sharply to zero in the $n$--$D$ phase map, a well-defined critical current, suppression by an out-of-plane magnetic field and by temperature, and oscillations of the critical current with out-of-plane magnetic field (Fig.~\ref{fig:si:zf1_characterization}).
Every session used Claude Opus 4.8.

\textsc{HERA} ran six sessions over six days (Table~\ref{tab:si:zf1_runs}).
After an initial test session, the recipe was amended to state the expected location of the superconducting region, which until then was given only in a paper the recipe cited~\cite{Zhang2023enhanced}.
However, the second session surveyed only up to $D/\epsilon_0 = -0.60$~V/nm, the largest $|D|$ at which the wide density linecuts the recipe suggests lie within the accessible gate range (Methods).
Because no linecut exhibited zero resistance, the session concluded that the device does not superconduct (Fig.~\ref{fig:main:transfer}c).
One operator message, pointing out that the accessible gate range had not been exhausted, was enough for \textsc{HERA} to sweep at higher $|D|$ and identify the superconductor (Fig.~\ref{fig:si:zf1_coverage}).
This failure led to the more important of the two corrections to the recipe, a suggestion to cover the whole accessible gate range before any negative conclusion.

With these two corrections, \textsc{HERA} found and characterized the superconductor autonomously, including the temperature dependence of the critical current (Fig.~\ref{fig:main:transfer}a,b).
The critical current, field, and temperature extracted by each of these sessions are within about a factor of two of those from a reference session that \textsc{HERA} ran with the operator in the loop (Figs.~\ref{fig:si:zf1_characterization}c and \ref{fig:si:zf1_gl}), with the spread set by where in the superconducting region each session chose to sit and how it extracted each quantity (Table~\ref{tab:si:zf1_runs}).

\section{Discussion}\label{sec:main:discussion}

We have demonstrated that an LLM agent can carry out a transport measurement campaign on quantum devices assembled from atomically thin materials from a brief to a defensible conclusion.
Given only the software tools of the corpus, both models in the stacking-order sessions planned reasonable measurements but took too little data to substantiate their claims.
The agent also needed the heuristics and rules of evidence of the corpus to draw conclusions that are both accurate and defensible. 
Even with the full corpus, sessions given the same brief varied in how they measured and in how well their data supported their conclusions, and expert grading made that variation measurable claim by claim. 
We therefore assessed the agent as one calibrates an instrument: the blinded trial, the ablation study, and the offline replay each measured an aspect of its behavior against a fixed reference.

Much of what the corpus holds is how to draw a conclusion, not how to perform one specific measurement.
A recipe in the corpus does not detail every feature a device will show, so the corpus also carries rules of evidence.
These rules require every claim to be made at a stated confidence level, alternative explanations to be considered where relevant, and a feature to count as absent only where the measurement could have resolved it.
Every tetralayer session at first misread the resistance peaks near charge neutrality, a feature no recipe covers (Sec.~\ref{sec:main:matrix}).
With the ablated corpus, this misreading led Opus 5 to rule out ABAB, while with the full corpus the rules of evidence kept every session from basing its stacking conclusion on the peaks.

The rules of evidence can make up for a gap in the corpus, but a mistake in the corpus affects every session that reads it.
That is why every change to the corpus is reviewed by a human and is kept under version control.
A human can only rule on a change if its effect can be measured, and measuring it with live sessions is costly.
Replay lets a change be tested offline and measured as a function of model at a much larger scale before it is used in the lab.
As of now, replay only tests a decision against previously acquired data.
A virtual lab, with simulated instruments and a forward model of the device, would let a recipe be tested from start to finish without cryostat time.

HERA is designed to be a hierarchy, with each layer operating at a different scope, and this work focused on the layer that runs a measurement campaign.
The layers above and below it are in development.
Below, a per-measurement layer that watches the data and optimizes measurements in real time would acquire high-quality data faster.
Above, a layer would direct a research program by drawing on theory and prior data to generate hypotheses for the campaign layer to test.
Assembled, the hierarchy would form a fully autonomous end-to-end discovery loop.

\section*{Acknowledgments}

The Claude family of models (Anthropic) was integral to the development of HERA, the measurements, and the writing of this manuscript.
We thank them and credit their contribution.
We thank Jacob Rebec, Junkai Dong, Slavko Rebec, Bingcheng Suo, Molly Andersen, Dan Parker, Ben Feldman, Marc Kastner, and David Goldhaber-Gordon for fruitful discussions.
This work was supported by an Anthropic AI for Science Program award.
Fabrication, measurement, and analysis performed by Q.C., S.S.K., B.H.A., Z.Z., and A.S. was supported by the Department of Energy, Office of Science, Basic Energy Sciences, Materials Sciences and Engineering Division, under Contract DE-AC02-76SF00515.
Fabrication, measurement, and analysis performed by J.T.M. were supported by the Department of Energy, Laboratory Directed Research and Development program at SLAC National Accelerator Laboratory, under contract DE-AC02-76SF00515.
S.S.K. acknowledges graduate financial support from the Knight-Hennessy fellowship.
K.W. and T.T. acknowledge support from the JSPS KAKENHI (Grant Numbers 21H05233 and 23H02052), the CREST (JPMJCR24A5), JST and World Premier International Research Center Initiative (WPI), MEXT, Japan.
Part of this work was performed at the Stanford Nano Shared Facilities (SNSF), supported by the National Science Foundation under Award No. ECCS-2026822.
Part of this work was performed at nano@stanford RRID:SCR\_026695.

\section*{Author Contributions}

Z.W.G., N.B., and A.S. conceived the project.
Z.W.G. developed HERA.
Z.W.G. and A.S. developed the recipes and corpus.
Q.C., J.M., B.H.A., and Z.Z. fabricated the devices.
K.W. and T.T. supplied the boron nitride crystals.
Z.W.G. and N.B. built the offline benchmark.
A.S. graded all sessions.
S.S.K. performed additional grading.
Z.W.G. analyzed the data.
Z.W.G. and A.S. wrote the manuscript with input from all authors.
A.S. supervised the project.

\section*{Competing interests}

This work was supported by an Anthropic AI for Science Program award.
Anthropic develops the Claude models evaluated in this work.
Anthropic had no role in the study design, data collection, analysis, or decision to publish.
N.B. is employed by Amazon Web Services, which has a commercial relationship with Anthropic.
This work was conducted independently of Amazon and without its support.

\section*{Data availability}

Every session archive (measurement notebooks, raw sweep data, device and rig cards, full agent transcripts), the pre-registration documents, the expert adjudications and grading, and the sealed documents are deposited in the Stanford Digital Repository at \url{https://purl.stanford.edu/tv195gf3182}.
The same snapshot is available at \url{https://github.com/sharpelab/hera-paper-data}.

\section*{Code availability}

The measurement harness, the agent prompts and corpus in the state used for each session, and the replay and metrics tooling are available at \url{https://github.com/sharpelab/hera-paper-data} and in the Stanford Digital Repository deposit at \url{https://purl.stanford.edu/tv195gf3182}.

\section{Methods}

\subsection*{Implementation}\label{sec:methods:implementation}

Each agent in the live sessions runs in Claude Code.
The live sessions ran Claude Opus and Fable models.
Each session draws on three documents.
A rig card describes the cryostat and its instruments.
A device card describes the device's wiring, gates, and safety limits.
A brief gives the session's goal and the operations it allows.
A setup agent assembles each session from the rig, device, brief, autonomy mode, and model the operator gives it.
It builds the session directory, fills the notebook's device-specific cells from the device card for the operator to confirm, and launches the measurement agent with the brief as its first message.
The measurement agent runs on the rig's control computer with shell access and no per-command permission prompt.
Instruments are driven through QCoDeS drivers~\cite{QCoDeS}, with drivers written by the lab for instruments QCoDeS does not cover.
Measurements utilize MeasureMe~\cite{MeasureMe}, which executes each sweep from a notebook cell, writes its data and metadata to disk, and plots it live.
The measurement agent writes to the operator's live Jupyter notebooks through a notebook channel.
The channel is a command-line tool that inserts cells into the notebook JupyterLab has open, rather than the file on disk, so the cells appear in the operator's open notebook as they are written.
The channel enforces a policy stored in each notebook.
The agent may append cells to the measurement notebook but cannot edit or delete them.
Whether the agent may also execute those cells is set by the session's autonomy mode.
In the analysis notebook, the agent may edit and execute cells.
Table~\ref{tab:si:models} gives the model, harness version, and corpus commit of every session and of each model in the replay.

\subsection*{Autonomy and operation}\label{sec:methods:operation}

The measurement agent runs in one of two autonomy modes, E0 and E1.
E0 is human-in-the-loop: the agent writes each measurement cell and the operator executes it.
E1 is autonomous: the agent executes the measurement cells it writes.
In both modes it executes its own analysis cells.
The operator first runs the setup and safety cells and then grants autonomy with a phrase the brief names.
The agent then runs a cell that reads the state of the rig without changing it and checks it against a list in the brief, such as the magnet being at zero field and the safety limits matching the device card.
The safety cell sets the device card's limits on the instrument parameters, so an out-of-range setting raises an error before any hardware moves.
The agent never changes a safety limit itself.
E1 engages only if every item on the list passes.
The brief also lists operations that need the operator's consent and operations that are banned.
If the agent has any concern for the safety of the device or the rig, it stops, notifies the operator, and returns to E0.
It resumes autonomous execution only when the operator grants autonomy again.
The operator can interrupt a running cell or withdraw the grant at any time.
On device 1 the agent stopped during its first gate leakage check, when the top-gate leakage rose superlinearly at voltages inside the safety limits of the device card.
The agent resumed after the operator ruled the leakage acceptable and granted autonomy again.

The agent messages the operator on Slack to report its progress and to notify the operator when it stops or needs them.
An intervention is any message the operator sends after granting autonomy, however minor.
The one exception is a message resent without new content because a provider outage interrupted the agent.
Every intervention, before and after each conclusion, is listed in Sec.~\ref{sec:si:interventions}.

Active time runs from the grant of autonomy to the agent's conclusion, excluding standby.
Standby is any stretch in which the agent is idle, waiting on the operator or on a provider outage.
Time in which a sweep is running counts as active.
Wall-clock time runs from the grant of autonomy to the end of the session and includes standby, the close-out, and any measurement the operator requested after the conclusion.

\subsection*{Corpus}\label{sec:methods:knowledge_base}

The corpus is kept in a git repository.
The same repository holds the agent prompts, the rig cards, and the templates for briefs, device cards, and notebooks.
When it was frozen for the trial, the corpus held 29 recipes totaling about 45\,000 words.
Five of the recipes are rig variants that state only how a base recipe changes on one rig.
A style guide sets how recipes are written.
Some statements in a recipe are rules, which say what an observation does or does not establish.
The guide requires each rule to state the conditions under which it holds, each load-bearing claim to cite its source and say what kind of source it is, and every conclusion to state its confidence level from a fixed list.
The five confidence levels are decisive, strong, weak, consistent-with, and uninformative.
The guide bars ``supports'', ``suggests'', and ``indicates'' from carrying the weight of a claim.
The guide sets no criterion for each level.
Instead, the recipes tie specific observations to specific levels.
The corpus also held 24 worked examples for 10 of the recipes, each a range of cells cropped from an earlier session notebook.
It also held 13 papers converted to text with their figures~\cite{Min2008chiral, Koshino2009gate, Zhang2010abc, Lui2011imaging, Cong2011raman, Koshino2011landau, Nguyen2014excitation, Grushina2015insulating, Datta2017, Zhou2021superconductivity, Zhang2023enhanced, Qin2026stripe, Kalantre2026fermiology}.
The corpus also held two forward models, which compute the Fermi surfaces, SdH frequencies, and cyclotron masses of the candidate stackings of trilayer and tetralayer graphene.
A further model extracts features from a Raman spectrum that distinguish stacking orders~\cite{Lui2011imaging, Cong2011raman}.

The software tools are the lab's analysis library and its documentation.
The library's functions convert raw data to physical quantities, for example gate voltages to $n$ and $D$.
The forward models and the Raman model are not counted among the software tools, and the ablated corpus removed them.

Every revision of the corpus is a commit.
When a session is built, its directory records the commit of the corpus it received and whether the corpus had uncommitted changes.
Sessions before 17 July 2026 predate this record.
For each of them, Table~\ref{tab:si:models} gives the last commit before the session began.
Every session of the blinded trial and of the corpus~$\times$~model study ran either on the frozen corpus, with nothing added but its device's brief and device card, or on the ablated corpus.

\subsection*{Review and revision}\label{sec:methods:revision}

After each session, the operator launches the review agent.
It archives the session and gathers findings from the session notes taken by the agent, the operator, and other observers.
It also reads the session transcript for problems, such as a rule the agent misread or an operation it repeatedly did by hand.
It also checks the session against every entry of the fix log.
The fix log is an append-only record of past revisions and deferred findings.
Each revision states what a later session should show if the revision worked, and each deferred finding states the evidence that would reopen it.
For each past revision, the agent checks whether the problem it addressed stayed fixed, was partly fixed, or recurred in the new session.
A problem that recurred, and a deferred finding whose evidence appeared, are added to the current session's findings.

For each finding, the agent recommends one of four actions: revise the corpus, defer the finding, only record it, or drop it.
For each recommended revision, it writes the suggested edit.
Entries to the fix log are committed without a ruling, and the operator can revert them.
The operator adjudicates each suggested revision.
Each accepted revision becomes a commit and an entry in the fix log.
The agent reports how many lines each revision adds or removes.
Every few sessions, a cleanup pass by the same agent merges and removes text without adding any, so that the corpus does not grow without limit.

\subsection*{Devices and fabrication}\label{sec:methods:devices}

Each device is hBN-encapsulated graphene, two to four layers thick, with graphite top and bottom gates.
The stacks were assembled by dry transfer with a polymer stamp, poly(bisphenol A carbonate) (PC) on polydimethylsiloxane (PDMS) for devices 1, 2, and 4 and polycaprolactone (PCL) on PDMS for device 3~\cite{Son2020strongly}. 
The stacks were deposited onto Si substrates with 285\,nm of SiO$_2$.
Each stack was patterned into a Hall bar by electron-beam lithography (Raith EBPG 5200+) and reactive ion etching (Oxford Plasma Pro 80), and Cr/Au contacts were deposited by electron-beam evaporation (Kurt J.\ Lesker LAB18).
The Hall bars of devices 1--3 were etched with CHF$_3$/O$_2$ and contacted at their edges~\cite{Wang2013one}.
For devices 1--3, the stacking order was identified by Raman spectroscopy at 532\,nm on a WITec alpha300 microscope~\cite{Cong2011raman, Nguyen2014excitation}.
For devices 1 and 2, the chosen region was isolated by ablation with a 1064\,nm laser on the same instrument.

The carrier density and displacement field follow from the two gate voltages as $n = (C_{t}V_{t} + C_{b}V_{b})/e$ and $D/\epsilon_0 = (C_{t}V_{t} - C_{b}V_{b})/2\epsilon_0$, where $C_{t}$ and $C_{b}$ are the top and bottom gate capacitances per unit area.
Each device card sets a safe range for each gate voltage.
Because $n$ and $D$ are linear in the two gate voltages, these ranges bound a parallelogram in the $n$--$D$ plane.

\textbf{Device 1: trilayer graphene device used in the blinded trial (Fig.~\ref{fig:si:devices}a).}
A rhombohedral region of an exfoliated trilayer flake was isolated.
A flat graphite flake was laser-cut to 20 by 25\,$\mu$m as the bottom gate.
The bottom hBN surface was tip-cleaned in contact mode on an AFM at 100\,nN before the top half of the stack was placed on it.
The top and bottom hBN were 14.9 and 21.4\,nm thick, and the Cr/Au contacts were 5/75\,nm thick.

\textbf{Device 2: tetralayer graphene device used in the blinded trial (Fig.~\ref{fig:si:devices}b).}
A rhombohedral region of a tetralayer flake was isolated.
The graphite top and bottom gates were laser-cut.
The crystal orientation was determined by atomically resolved torsional force microscopy on a Park NX-HiVac with an Opus 240AC-SG cantilever.
The top and bottom hBN were 9.5 and 11\,nm thick, and the Cr/Au contacts were 5/75\,nm thick.
The rhombohedral stacking likely relaxed to Bernal during stacking.

\textbf{Device 3: trilayer graphene device used in the development campaign (Fig.~\ref{fig:si:devices}c).}
A Bernal trilayer flake, confirmed by Raman spectroscopy~\cite{Zhou2021half}, was picked up with the advancing contact front along its armchair axis in an attempt to convert it to rhombohedral stacking.
The bottom hBN surface was protected from polymer by a sacrificial hBN flake that was slid away before the top half was placed~\cite{Pack2025sliding}, and the crystal orientation was determined by torsional force microscopy on a Bruker Icon AFM~\cite{Pendharkar2024torsional}.
The top and bottom hBN were each 10--20\,nm thick, and the Cr/Au contacts were 5/90\,nm thick.
After fabrication, Raman showed rhombohedral stacking in the ungated regions near the contacts, but the graphite gates prevented Raman from reaching the channel.

\textbf{Device 4: Bernal bilayer graphene/WSe$_2$ device used in the search for superconductivity (Fig.~\ref{fig:si:devices}d).}
The top half of the stack held hBN, a Bernal bilayer graphene flake, and few-layer WSe$_2$, picked up in that order.
Both halves of the stack were annealed in Ar/H$_2$ at 350\,\textdegree C for 12\,h and tip-cleaned in contact mode on a Bruker Dimension IconIR AFM.
The top and bottom hBN were 23.6 and 20\,nm thick, and the Cr/Au top contacts were 5/65\,nm thick.

\subsection*{Rigs}\label{sec:methods:cryostats}
The three multilayer graphene devices were measured in a variable temperature insert with a base temperature of about 1.2~K.
The measurement lines of the variable temperature insert have no filtering.
The WSe$_2$-proximitized bilayer graphene device was measured in a Leiden Cryogenics CF-900 dilution refrigerator with a base temperature of about 30~mK.
The probe used in the dilution refrigerator was a two-axis rotation probe with attocube rotation stages.
The probe consists of coaxial wiring with custom filtering.
The wires first pass through a cured mixture of epoxy and bronze powder to filter GHz frequencies, then through low-pass RC filters mounted on sapphire plates to filter MHz frequencies.
In both cryostats, Stanford Research Systems SR830 and SR860 lock-in amplifiers were used to perform four-terminal resistance measurements.
On the dilution refrigerator, the lock-ins were paired with NF Corporation LI-75A or Basel Precision Instruments SP1004 voltage preamplifiers.
The current was measured by a DL Instruments 1211 current preamplifier.
Keithley 2400 source-measure units and Stanford Research Systems DC205 voltage sources applied the gate voltages.

\subsection*{Blinded trial protocol}\label{sec:methods:trial}

The protocol for the trial was committed to the repository that holds the corpus on 17 July 2026.
The corpus, the templates for the brief and device card, and the agent prompts were frozen under a git tag on 19 August 2026, before the first trial session.
The protocol's endpoint was a conclusion on the stacking order for each measured contact pair in the form of the stacking recipe's output contract.
The stacking recipes end in an output contract that gives a confidence level to every per-region verdict.
The contract pairs each verdict with its evidence, the named hedges, and the measurement that would strengthen the conclusion.
Under the contract, a ``cannot tell'' with reasons is a valid outcome.
A confident conclusion without feature-level evidence is a protocol violation.
The protocol required that all results of every eligible device be published in the order they were measured.
Every departure from the protocol was logged with its date and is published with the data.

Before each session, the expert sent the operator two SHA-256 hashes.
One was the hash of his conclusion on the device's stacking order and the associated confidence level.
His confidence level used the same scale as the output contract.
The other was the hash of the device's Raman record.
The operator committed both hashes to the repository, which fixed the two documents before the session without revealing them.
For device 2 the Raman record was the expert's written reading of the parent flake, because the raw map could not be retrieved before the transport measurements started.

The agent received only the brief, the device and rig cards, and the corpus.
It received no earlier measurement of the device, no Raman, and no sign that the session was part of a trial.
The expert assigned each device its alias and supplied the facts its cards were built from.
Before each session, the brief, cards, and notebooks were searched for words that would tell the agent its result would be compared with a known answer, such as ``benchmark'', ``ground truth'', or ``replication''.
The operator saw neither the Raman nor the expert's conclusion until the agent's conclusion was committed to the repository.
The Raman record was then committed and checked against its hash. 
A fresh instance of the agent was given the committed conclusion, the Raman record, and the session's notes and analysis notebook, and wrote a fused verdict.
The protocol named the fused verdict as a secondary endpoint.

Only then was the expert's conclusion opened, checked against its hash, and compared with the agent's conclusion.
The comparison set the agent's conclusion on each contact pair beside the expert's.
It recorded whether the two agreed in stacking order and in confidence level.
The protocol reports this agreement rather than accuracy.
The protocol named a separate adjudication by the expert, written after reading the agent's conclusion, as a secondary endpoint, and his sealed conclusion was used in its place.

\subsection*{Corpus $\times$ model protocol}\label{sec:methods:matrix}

The corpus~$\times$~model study ran Fable~5 and Opus~5, each once on the full corpus and once on the ablated corpus.
The four corpus~$\times$~model sessions ran after the trial and outside its protocol, on the same cooldown of device 2, with the same operator.
The trial session on device 2 also ran Fable~5 on the full corpus.
That cell of the matrix therefore holds two sessions, and every other cell holds one.
The later of the two is called the Fable~5 repeat.
The ablated corpus removed the recipes, worked examples, forward models, papers, style guide, and method documents.
It kept the cards, the software tools with their reference, and the safety limits.
References to the removed material were also edited out of the agent prompt, the cards, and the tool reference.
The full-corpus sessions ran on one brief, and the ablated-corpus sessions on a second.
The second brief had the same goal, limits, and contacts.
It asked the agent to derive a measurement plan from the cards and the tool reference and to write the plan down before its first measurement.
Because the ablated corpus had no style guide, the second brief also carried its own output contract, with the confidence levels weak, moderate, and strong.

\subsection*{Grading}\label{sec:methods:grading}

An LLM (Sonnet~5) extracted the claims from each session's transcript, assistant messages only, and from the markdown of its notebooks, under a fixed prompt.
A claim is any statement or choice a careful experimentalist could disagree with.
Each claim was categorized as an inference, a decision, or a verdict.
Observations were not graded on their own but were folded into the inferences they support.
Claims about gate leakage and claims that a setting lay within the card's limits were excluded.
The extraction rules were calibrated on one session that was not graded.
Each claim was split into checkable items.
The claims, their items, and the pre-registration were locked before any item was graded (tag \texttt{t1-scoring-locked} of the grading repository).
Device names, dates, session names, corpus hashes, and people's names were removed from every claim before the claims were pooled.
The grader could flag any claim whose session he recognized.
He flagged none of his 151 claims as recognized, and the second expert flagged none of his 56.

Each item was scored 1 if correct, 0.5 if partly correct, and 0 if wrong, and unverifiable items were excluded. 
Item scores were averaged to give each claim's grade.
Claim grades were averaged by type to give each session's inference, decision, and verdict grades.
All of a session's claim grades, with verdict claims weighted double, were averaged to give its claim score.
The claim-by-claim grading took the expert about 8 to 12 hours over two days.

A second expert (S.S.K.) graded a fixed subset of the claims with the same rubric.
He had not run any session or seen any session's notebooks.
The subset held every verdict claim and 30\% of the inference and decision claims of each session.
In all it held 56 claims and 212 items, fixed and shuffled before they were delivered to him.
Before grading, he knew that trilayer devices had been measured with rhombohedral stacking as the aim, and he had attended a talk by J.M. on measurements of device 1.
He did not know which sessions used which model or corpus.
He took about two hours over eight sittings.
The first expert's grades are the grades of record.
The second expert's grades are reported beside them and are never averaged in.
Agreement was measured on the items both experts graded correct, partly correct, or wrong, as exact agreement and as quadratic-weighted Cohen's $\kappa$.
Both measures, the script that computes them, and the strata reported in Supplementary Sec.~\ref{sec:si:second_expert} were fixed before the claims were delivered to him.

The expert who graded every claim also read each session as a whole and gave it two grades.
The first grade asked whether the session answered the brief, with options A (as it stands), B (after specified re-analysis), C (only after re-measurement), and F (not at all).
The second asked how much supervision he would give the agent next time, with options A (hand it the cryostat and read the draft), B (trust its sweeps and check its analysis), C (check every decision), and F (not alone with the rig).
He graded the sessions in a random order fixed in advance.
This grading was not blind to the device, because he recognized it from its gate maps.

\subsection*{Replay benchmark}\label{sec:methods:replay}

Each replay started a fresh agent from the archived state of the Fable~5 repeat, taken after both SdH spot checks had been acquired and before the fan was designed.
The archived state held the session's notebooks, data, and notes, but not the live transcript.
The live agent's analysis of the first spot check and its draft of the fan were removed.
So were its notes beside the spot-check cells on the field step and the lowest field the fan would need.
A test confirmed that none of the removed material remained.
Each replay ran in a container holding the archived state.
The Claude models ran in Claude Code, and GPT-6 Astra and GPT-6 Sol ran in OpenAI's Codex command-line harness.
The agent's brief ended with a request for a Landau fan, and the replay ended at the first measurement cell the agent wrote.
Each model ran at high effort until it had designed twenty fans with each corpus, with at most forty replays per corpus.
Only replays that ended in a fan were graded.
With the full corpus, GPT-6 Astra designed only nineteen fans, because it spent 21 of its 40 replays on spot checks.

A deterministic grader scored each fan against the expected SdH frequencies for tetralayer graphene.
Rhombohedral stacking is expected to exhibit one SdH frequency, and Bernal stacking two.
The two Bernal frequencies were calculated using a published model of Bernal tetralayer graphene~\cite{Halder2026tetralayer}.
A fan can discriminate between the two stacking orders if it resolves the light Bernal frequency and separates the rhombohedral frequency from the heavy Bernal frequency.
To do so, the designed fan's field step size must sample at least two points per period of the fastest frequency.
Its record length in $1/B$, counting only the field steps that meet this condition, must also be long enough to resolve the slowest frequency.

Four metrics were pre-registered on 29 September 2026, before any production replay ran: $P(\text{call})$, frequency coverage, sampling rate, and Pareto dominance.
The first, $P(\text{call})$, is the probability that the fan can discriminate between the two stacking orders.
To compute it, the grader scored each constant-density linecut of the fan by how well it resolved the light Bernal frequency and how well it separated the rhombohedral frequency from the heavy Bernal frequency.
The record started at the onset field of each oscillation on device 2.
The onset field was 0.75~T for the light Bernal frequency and 1.0~T for the rhombohedral and heavy Bernal frequencies.
Each score rose from zero to one as the record length grew from 1.5 to 3.5 periods of the light Bernal frequency.
The grader also read the instrument settings of the fan.
A linecut's score was reduced if the excitation current and lock-in time constant were too small to detect the frequencies, and set to zero if the expected signal would overload the lock-in sensitivity.
Only linecuts on the electron side counted.
On the hole side, the safe gate range did not reach densities at which the light Bernal frequency could be resolved.
Two of the 239 scored fans, both from GPT-6 Astra with the full corpus, covered only the hole side and scored zero; excluding them changes the pooled gain by 0.01.

The grader treated each linecut's score as the probability that its Fourier transform would show the frequencies.
Two linecuts at nearly the same density give nearly the same frequencies, so linecuts counted separately only if their expected light Bernal frequencies differed by at least half an FFT bin.
A single detected linecut shows the frequencies at one density, while the call rests on how they change with density.
The grader therefore set the probability of making the call to rise with the number of detected linecuts, with one, two, three, and four or more giving 0.16, 0.50, 0.84, and 1, respectively.
$P(\text{call})$ is this probability averaged over the possible numbers of detected linecuts.
This was computed separately for resolving the light Bernal frequency and for separating the rhombohedral frequency from the heavy Bernal frequency.
The fan's $P(\text{call})$ was the smaller of the two results.
The pre-registered metric was $P(\text{call})$ with a cutoff of eight hours, and a fan that would take longer to acquire scored zero.
The two fans HERA measured on device 2 that the grader scored at $P(\text{call}) = 1$ took 3.5 and 4.6~h, so the cutoff is generous.

Frequency coverage is the range of light Bernal frequency that the fan's linecuts cover, in bins of the fan's Fourier transform.
Each linecut covered half a bin centred on its expected light Bernal frequency, weighted by its score.
Sampling rate is the number of field points per period of the rhombohedral frequency at the fan's highest electron-side density.
It was read over the fan's coarsest field step above 1.0~T.
Pareto dominance compared every pair of fans from the same model, one designed with each corpus.
One fan dominated the other if it was no worse on both $P(\text{call})$ and acquisition time and better on at least one.
Pareto dominance is the fraction of pairs in which the full-corpus fan dominated, minus the fraction in which the ablated-corpus fan dominated, averaged over models.
Acquisition time was estimated from a timing model of the rig, which reproduces the durations of the live fans to within 8\%.

Each metric was compared between the two corpora by a two-sided permutation test, with the corpus labels shuffled 10\,000 times within each model.
The four $p$ values were Holm-adjusted for multiple comparisons.
Cutoffs of six hours, twelve hours, and none were pre-registered as sensitivity analyses (Table~\ref{tab:si:budget}).
Every model's gain was positive at every cutoff except that of Fable~5 at twelve hours, which was $-0.02$.
Fable~5's gain depends on the cutoff because eleven of its twenty full-corpus fans would take 9.4 to 20.4~h, while its best ablated-corpus fans take 7 to 9~h.
The other pre-registered sensitivity analyses varied the grader's settings or scored replays without a fan as zero.
In each, the full corpus still raised the pooled $P(\text{call})$, by 0.23 to 0.45.

The grader was developed on all replays run before it was frozen.
The range of 1.5 to 3.5 periods over which each score rises was set by how often the frequencies were detected in three fans \textsc{HERA} measured on device 2.
The reduction for a small current and time constant was calibrated on the signal-to-noise ratio measured in two of these fans, at 50 and 10~nA.
The four metrics were chosen from about twenty candidates by how well they separated the two corpora on the development fans.
The development fans were 27 designed in replays from the same state with no request for a fan, and the 5 HERA measured on device 2.
Fifteen of the 19 development fans designed with the full corpus came from models other than the six tested here.
The full corpus suggests that the agent sample the field axis uniformly in $1/B$, a design the metrics tend to favor.
The 21 ablated-corpus fans that used such a grid scored as well as the full-corpus fans, with a mean $P(\text{call})$ of 0.97.
The starting state included 3.6~h of spot checks that the full corpus had prompted.
Replays with either corpus therefore started with these spot checks already measured, so the gap between the corpora offline is a lower bound on the gap in live sessions.
$P(\text{call})$ treats detection failures in different linecuts as independent.
In a fan HERA measured at 10~nA, the failures clustered in density, so $P(\text{call})$ overestimates fans at low current.

\subsection*{Cost}\label{sec:methods:cost}

Tokens were counted once per API response, although the transcript repeats a response's usage on each content block it records.
Tokens were priced by class, as uncached input, output, cache reads, and cache writes.
The live sessions wrote to the one-hour cache.
At list price, uncached input and output cost \$10 and \$50 per million tokens for Fable 5, and \$5 and \$25 for Opus 5.
Cached input costs a tenth of the input price.
Writing to the cache costs 1.25 times the input price for the five-minute cache and twice the input price for the one-hour cache.

\bibliography{main}

\clearpage
\renewcommand\thefigure{S\arabic{figure}}
\setcounter{figure}{0}
\renewcommand\thetable{S\arabic{table}}
\setcounter{table}{0}
\renewcommand\thesubsection{S\arabic{subsection}}
\setcounter{subsection}{0}

\clearpage
\title{Supplementary Information for HERA: Assessing an Autonomous Agent for Quantum Device Measurement}
\maketitle
\beginsupplement
\onecolumngrid
\makeatletter
\pretocmd{\section}{\clearpage}{}{}
\makeatother
\startcontents[SI]
\printcontents[SI]{}{1}{\setcounter{tocdepth}{2}}

\FloatBarrier

\section{Devices}\label{sec:si:devices}

\begin{figure}[htbp]
\input{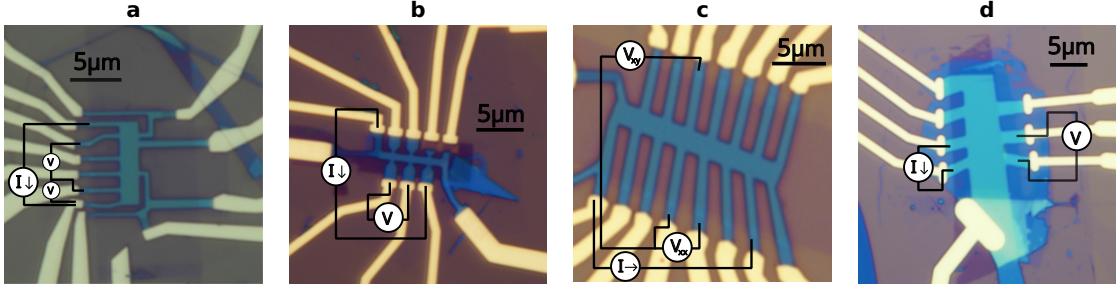}
\caption{\label{fig:si:devices}
\textbf{Optical micrographs of the four devices.}
The current path (I) and voltage contacts (V) used by HERA are drawn on each image.
(\textbf{a}) Device 1, the trilayer graphene device used in the blinded trial.
(\textbf{b}) Device 2, the tetralayer graphene device used in the blinded trial.
(\textbf{c}) Device 3, the trilayer graphene device used in the development campaign.
(\textbf{d}) Device 4, the Bernal bilayer graphene/WSe$_2$ device used in the search for superconductivity.
Scale bars, 5~$\mu$m.}
\end{figure}

\FloatBarrier

\section{Models and versions}\label{sec:si:models}

\begin{table}[htbp]
\small
\begin{ruledtabular}
\begin{tabular}{lllclcc}
Experiment & Model & Model id & Effort & Harness & Corpus & Mode \\
\colrule
\multicolumn{7}{l}{\textit{Blinded trial}} \\
Device 1 & Fable 5 & \texttt{claude-fable-5} & high & CC 2.1.235 & \texttt{d8e1337} & E1 \\
Device 2 & Fable 5 & \texttt{claude-fable-5} & high & CC 2.1.239 & \texttt{13b07e9} & E1 \\
\colrule
\multicolumn{7}{l}{\textit{Corpus $\times$ model, device 2}} \\
Full corpus, repeat & Fable 5 & \texttt{claude-fable-5} & high & CC 2.1.241 & \texttt{13b07e9} & E1 \\
Full corpus & Opus 5 & \texttt{claude-opus-5} & high & CC 2.1.243 & \texttt{13b07e9} & E1 \\
Ablated corpus & Fable 5 & \texttt{claude-fable-5} & high & CC 2.1.241 & \texttt{039b0f5} & E1 \\
Ablated corpus & Opus 5 & \texttt{claude-opus-5} & high & CC 2.1.241 & \texttt{039b0f5} & E1 \\
\colrule
\multicolumn{7}{l}{\textit{Offline replay, number of replays with the full + ablated corpus}} \\
36 + 20 & Fable 5 & \texttt{claude-fable-5} & high & CC 2.1.283 & \texttt{13b07e9}\,/\,\texttt{039b0f5} & n/a \\
33 + 20 & Fable 5.1 & \texttt{claude-fable-5-1} & high & CC 2.1.283 & \texttt{13b07e9}\,/\,\texttt{039b0f5} & n/a \\
24 + 21 & Opus 5 & \texttt{claude-opus-5} & high & CC 2.1.283 & \texttt{13b07e9}\,/\,\texttt{039b0f5} & n/a \\
25 + 20 & Opus 5.5 & \texttt{claude-opus-5-5} & high & CC 2.1.283 & \texttt{13b07e9}\,/\,\texttt{039b0f5} & n/a \\
40 + 20 & GPT-6 Astra & \texttt{gpt-6-astra} & high & Codex 0.157.1 & \texttt{13b07e9}\,/\,\texttt{039b0f5} & n/a \\
20 + 21 & GPT-6 Sol & \texttt{gpt-6-sol} & high & Codex 0.157.1 & \texttt{13b07e9}\,/\,\texttt{039b0f5} & n/a \\
\colrule
\multicolumn{7}{l}{\textit{Domain transfer, WSe$_2$-proximitized bilayer graphene}} \\
Session 1 & Opus 4.8 & \texttt{claude-opus-4-8} & --- & CC 2.1.183 & \texttt{d3b7885} & E0 \\
Session 2 & Opus 4.8 & \texttt{claude-opus-4-8} & --- & CC 2.1.183 & \texttt{12fa296} & E0 \\
Session 3 & Opus 4.8 & \texttt{claude-opus-4-8} & --- & CC 2.1.186 & \texttt{f1fc725} & E0 \\
Session 4 & --- & --- & --- & CC 2.1.187 & \texttt{84857f8} & E0 \\
Session 5 & Opus 4.8 & --- & --- & CC 2.1.191 & \texttt{256ccd5} & E1 \\
Session 6 & Opus 4.8 & \texttt{claude-opus-4-8} & --- & CC 2.1.191 & \texttt{591da57} & E1 \\
Reference session & Opus 4.8 & \texttt{claude-opus-4-8} & --- & CC 2.1.191 & \texttt{f87396c} & E0 \\
\colrule
\multicolumn{7}{l}{\textit{Development campaign, device 3, numbered by run}} \\
Session 1 & Fable 5 & \texttt{claude-fable-5} & --- & CC 2.1.170 & \texttt{d197ec7} & E0 \\
Session 2 & Opus 4.8 & \texttt{claude-opus-4-8} & --- & CC 2.1.177 & \texttt{78afaf6} & E0 \\
Session 3 & Opus 4.8 & \texttt{claude-opus-4-8} & --- & CC 2.1.178 & \texttt{5a18746} & E1 \\
Session 4 & Opus 4.8 & \texttt{claude-opus-4-8} & --- & CC 2.1.179 & \texttt{6a3fa74} & E1 \\
Session 5 & Opus 4.8 & \texttt{claude-opus-4-8} & --- & CC 2.1.183 & \texttt{9f72450} & E1 \\
Session 7 & Opus 4.8 & \texttt{claude-opus-4-8} & --- & CC 2.1.183 & \texttt{425bc8d} & E1 \\
Session 8 & Opus 4.8 & \texttt{claude-opus-4-8} & --- & CC 2.1.183 & \texttt{6c810ff} & E1 \\
Session 9 & Opus 4.8 & \texttt{claude-opus-4-8} & --- & CC 2.1.183 & \texttt{c54361c} & E1 \\
Session 10 & Opus 4.8 & \texttt{claude-opus-4-8} & --- & CC 2.1.183 & \texttt{870c29d} & E1 \\
\end{tabular}
\end{ruledtabular}
\caption{\label{tab:si:models}\textbf{Models, harness, and corpus of each experiment.}
One row per live session, and one per model for the offline replay.
Model id is the model field of the session transcript, or of the replay case file and every replay transcript.
Effort is the reasoning-effort setting the harness recorded; the harness versions of June 2026 record none, though their transcripts contain the model's thinking.
Harness is the Claude Code (CC) or Codex CLI version.
Corpus is the commit of the lab repository the session ran on: \texttt{d8e1337} is the corpus frozen for the trial, \texttt{13b07e9} the same corpus with the brief and device card for device 2 added, and \texttt{039b0f5} the ablated corpus.
Sessions before 17 July 2026 did not record their commit, so for the transfer and development sessions the commit given is the last in the repository before the session's transcript begins.
Mode is the autonomy mode (Methods); n/a marks the offline replay, and a dash a value that was not recorded.
The conversation record of transfer session 4 was lost, so its model is not recorded; the setup agent that launched it ran \texttt{claude-opus-4-8}.
Transfer session 5 survives as a terminal export that names Opus 4.8 but not the model id.
The reference session survives as a terminal export and the harness's own record, which name its model.
Run 6 of the development campaign, a false start, is omitted (Fig.~\ref{fig:si:ladder}).}
\end{table}

\FloatBarrier

\section{Determining stacking order from a Landau fan}\label{sec:si:stacking}

A standard way to determine stacking order from transport is a Landau fan.
A Landau fan is the resistance measured as a function of magnetic field $B$ and carrier density $n$.
The fan can be interpreted in two complementary ways.
One is to Fourier transform the Shubnikov--de Haas (SdH) oscillations at each density and count the frequencies in the spectrum.
SdH oscillations are oscillations of the resistance that are periodic in $1/B$.
Each extremal orbit on the Fermi surface contributes one frequency, proportional to the area the orbit encloses~\cite{Onsager1952}.
At sufficiently high density, rhombohedral stacking has one such orbit and Bernal stacking more than one, because their band structures differ.
The measured frequencies can therefore be followed across density and compared with those computed from the single-particle band structure of each stacking order~\cite{Min2008chiral, Koshino2009gate, Zhang2010abc}.
The other is to follow the sequence of quantum Hall gaps, each a straight line in the $(n, B)$ plane whose slope gives its filling factor~\cite{Streda1982}.
In a tetralayer, Bernal and rhombohedral stacking have the same sequence of filling factors at low field~\cite{Min2008chiral} and differ only in the sizes of the gaps and the crossings between levels~\cite{Koshino2011landau}.

In a trilayer at the densities measured here, ABC stacking has one extremal orbit, whose frequency $h|n|/4e$ is set by the density alone~\cite{Zhang2010abc}.
ABA stacking has two, splitting the carriers between a light, monolayer-like band and a heavier, bilayer-like band~\cite{Koshino2009gate}.
The two ABA frequencies sum to the ABC frequency.
The heavier band holds most of the carriers, so one frequency of the ABA pair lies close to the ABC frequency.
The corpus supplies a single-particle model with the ABA band parameters of Ref.~\cite{Datta2017}.
At $n = -1.74\times10^{12}$~cm$^{-2}$, it places the ABA pair at 2.3 and 15.7~T and the ABC frequency at 18.0~T.
The larger ABA frequency is thus 0.87 of the ABC frequency.
A density estimated from the gate capacitances, using the measured hBN thicknesses, is uncertain by about 10\%.
A single frequency near the ABC frequency is therefore difficult to assign, because the ABC frequency and the heavier ABA frequency differ by only 13\% at this density.
The light-band frequency therefore separates the two stacking orders more clearly than the heavier one.
In a tetralayer the picture is the same~\cite{Halder2026tetralayer}.
ABCA stacking has one frequency, $h|n|/4e$.
ABAB stacking has two, from a light band and a heavy band.
The two ABAB frequencies sum to the ABCA frequency.
Unlike the tetralayer's, the quantum Hall gap sequences of the two trilayer stacking orders differ at low field.
ABA stacking has gaps at every even filling factor~\cite{Taychatanapat2011}, and ABC stacking at every fourth filling factor plus two, $\nu = \pm 6, \pm 10, \pm 14, \ldots$~\cite{Bao2011stacking}.

Whether a Landau fan can tell the stacking orders apart depends on its design.
The design sets the range of density and field the fan covers and how finely both axes are sampled.
The range and sampling a fan needs are set by the onset field of the oscillations and by the highest frequency to be resolved.
A fan that meets these requirements also resolves the quantum Hall gaps.
Locating a crossing between levels requires fine steps in density.
The corpus holds a recipe for determining stacking order that covers fan design.
The recipe treats the frequency count as the primary discriminator and any resolved crossing as a complement.
The recipe suggests short SdH spot checks at a few densities.
It also suggests checks for artifacts in the SdH spectra, such as harmonics and magnetic breakdown~\cite{Leeb2025fieldguide}, before any frequency is counted.
The recipe asks for a hedged conclusion that cites its evidence.
It counts a reasoned ``cannot tell'' as a successful outcome.

\begin{figure}[htbp]
\centering
\input{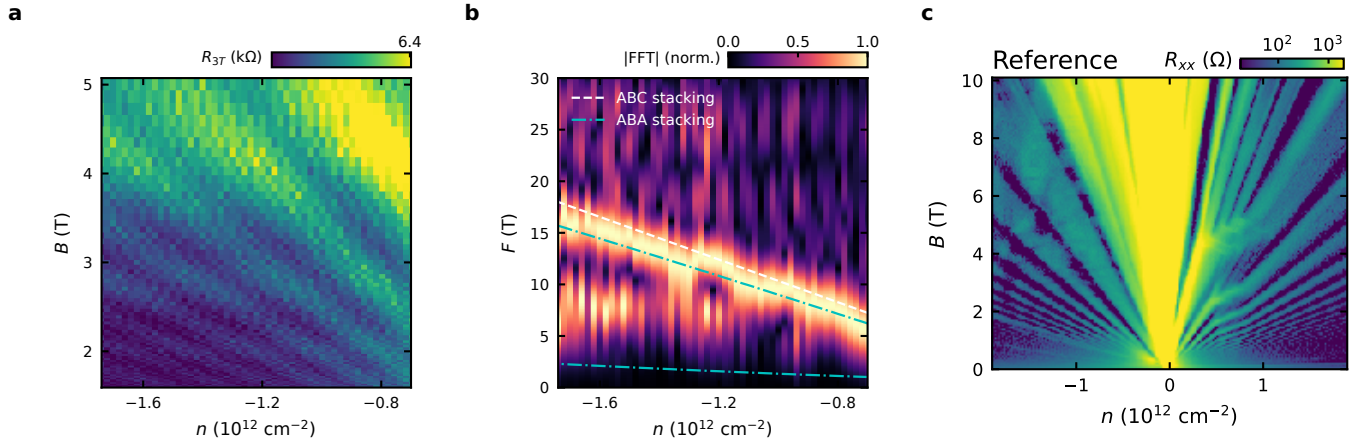}
\caption{\label{fig:si:trial_fans}
\textbf{The Landau fans behind the trilayer conclusions.}
(\textbf{a}) The fan behind HERA's stacking-order conclusion.
Three-terminal resistance $R_{3T}$ as a function of $n$ and $B$ at $D = 0$, from 1.6 to 5~T uniform in $1/B$, acquired in 1.7~h at 1~nA with a 100~ms time constant.
(\textbf{b}) Fourier amplitude $|\mathrm{FFT}|$ of each constant-density linecut of (\textbf{a}), each normalized to its maximum, as a function of $n$ and SdH frequency $F$.
The lines indicate the frequencies expected for ABC and ABA stacking at the density calibration HERA set in the session.
(\textbf{c}) The fan behind the expert's sealed conclusion.
Four-terminal longitudinal resistance $R_{xx}$ as a function of $n$ and $B$, measured on the same device and cooldown before HERA's session.
}
\end{figure}

The range and sampling requirements on a fan can be stated as numbers.
SdH oscillations are periodic in $1/B$, so the frequency resolution of a record is the inverse of its range in $1/B$, $\delta F = 1/(1/B_{\min} - 1/B_{\max})$.
The range must hold several periods of the lowest frequency to be resolved.
The step in $1/B$ must be below half the period of the highest frequency, $\Delta(1/B) < 1/(2F_{\max})$.
Otherwise that frequency aliases into the low-frequency band, where it can pass for the light-band frequency.
Because the ABC frequency is the highest frequency expected, the recipe sets the step size from it.
These are the two requirements the replay grader scores (Methods).
The trilayer fan on which HERA based its conclusion spans 1.6 to 5~T (Fig.~\ref{fig:si:trial_fans}a).
Over this range the light ABA frequency completes about one period.

A spot check is a field sweep at fixed density and displacement field.
It measures the onset field of the oscillations and the frequencies present at that density, which set the field range and step of the fan.
On the variable temperature insert, spot checks are sampled uniformly in $B$.
The step is chosen so that the sampling condition above holds at the lowest field.
In the trilayer session, HERA ran two spot checks at $D = 0$, one at $n = -1.8$ and one at $+1.7\times10^{12}$~cm$^{-2}$.
Each swept from 0.8 to 9~T uniformly in $B$ in 11~mT steps, 746 points in 1.9~h.
At 0.8~T this step is $\Delta(1/B) = 0.017$~T$^{-1}$, three points per period of the ABC frequency at the hole density.
The hole-side check found an onset field of the SdH oscillations near 1.7~T and two frequencies that held across the background and window choices.
The electron-side check found an onset field near 2.2~T, with its low-frequency region sensitive to the background.
HERA therefore planned the fan on the hole side.

The amplitude of each SdH frequency falls with temperature as $X/\sinh X$, where $X = 2\pi^2 k_B T m^*/(\hbar e B)$ and $m^*$ is the cyclotron mass of the carriers on that orbit~\cite{Lifshitz1956}.
A temperature series of the oscillations therefore gives a mass that can be compared with the band structure of each stacking order.
In a trilayer the mass separates the two readings of a single frequency near the ABC frequency, because the heavier ABA band and the ABC band have different masses.
HERA took a temperature series at $n = -1.74\times10^{12}$~cm$^{-2}$ and $D = 0$, sweeping the field from 2 to 5~T at seven temperatures from 1.25 to 15.1~K (runs 12--18).
It fitted the Fourier amplitude of the dominant frequency, near 16~T, to the Lifshitz--Kosevich form at $B = 2.86$~T.
The fit gave $m^* = 0.049 \pm 0.002\,m_e$.
The single-particle model in the corpus gives $0.060\,m_e$ for the heavier ABA band and $0.11\,m_e$ for ABC at this density.
HERA read the 18\% difference from the ABA model as within the model's parameter uncertainty.
It excluded ABC.
The mass of the light ABA band, $0.010\,m_e$ in the model, was not measured.
The fit takes the sample thermometer as the temperature, while the rig card puts the electron temperature near 8~K at base.
Refitting without the two temperatures below 4~K gives $0.051 \pm 0.003\,m_e$.

At zero field, the tetralayer device shows two resistance peaks near charge neutrality, with a dip between them.
A Bernal tetralayer is expected to show more than one such peak, because it has Lifshitz transitions near neutrality~\cite{Shi2018lifshitz, Hirahara2018peaks}.
Under a displacement field, the resistance at neutrality rises and saturates, which indicates a gap of a few meV~\cite{Shi2018lifshitz}.
The stacking recipe does not name this feature.
All five sessions on the device first read the two peaks as two regions of different doping along the channel, or as a segment gated by only one gate.
The sessions with the full corpus did not rest their stacking conclusions on that reading.
The Opus~5 session with the ablated corpus took a ten-fold rise in resistance under displacement field, at the peak it took for neutrality, as evidence against Bernal stacking.
It excluded ABAB partly on that evidence.
The expert graded this reasoning wrong.

\FloatBarrier

\section{The six live stacking-order sessions}\label{sec:si:trial_table}

\begin{table}[htbp]
\centering
\small
\setlength{\tabcolsep}{4pt}
\begin{ruledtabular}
\begin{tabular}{lcccccc}
 & \multicolumn{2}{c}{Blinded trial} & \multicolumn{4}{c}{Corpus $\times$ model, device 2} \\
\cmidrule(lr){2-3}\cmidrule(lr){4-7}
Device (layers) & 1 (3L) & 2 (4L) & 2 (4L) & 2 (4L) & 2 (4L) & 2 (4L) \\
Model & Fable 5 & Fable 5 & Fable 5 & Opus 5 & Fable 5 & Opus 5 \\
Corpus & full & full & full, repeat & full & ablated & ablated \\
\colrule
\multicolumn{7}{l}{\textit{Session}} \\
Start (2026) & Aug 18 & Aug 21 & Aug 22 & Aug 24 & Aug 23 & Aug 24 \\
Sweeps, before $+$ after conclusion & 19 + 2 & 16 + 0 & 15 + 0 & 8 + 1 & 19 + 0 & 13 + 0 \\
\colrule
\multicolumn{7}{l}{\textit{Conclusion}} \\
Stacking order & ABA & ABAB & ABAB & ABAB & ABAB & ABCA \\
Confidence & strong & strong & strong & consistent-with & strong & moderate \\
Expert agrees & yes & yes & yes & order only & yes & no \\
\colrule
\multicolumn{7}{l}{\textit{Operator}} \\
Interventions before conclusion & 1 & 0 & 0 & 0 & 0 & 1 \\
Interventions after conclusion & 1 & 0 & 0 & 2 & 1 & 2 \\
Standby (h) & 16.6 & 0 & 0 & 0 & 10.9 & 0.4 \\
Sweep time after conclusion (h) & 8.1 & 0 & 0 & 4.6 & 0 & 0 \\
\colrule
\multicolumn{7}{l}{\textit{Time and cost}} \\
Active (h) & 16.2 & 12.9 & 14.4 & 5.5 & 3.9 & 4.1 \\
Wall (h) & 42.1 & 13.3 & 15.3 & 11.9 & 14.8 & 4.7 \\
Cost (\$) & 265 & 196 & 200 & 146 & 119 & 100 \\
\colrule
\multicolumn{7}{l}{\textit{Expert grading}} \\
Claims ($n$) & 29 & 28 & 25 & 24 & 17 & 28 \\
Items ($n$) & 122 & 104 & 94 & 85 & 75 & 81 \\
Inference & 0.86 & 0.68 & 0.91 & 0.74 & 0.44 & 0.58 \\
Decision & 0.95 & 0.95 & 0.96 & 0.85 & 0.71 & 0.93 \\
Verdict & 1.00 & 1.00 & 0.43 & 1.00 & 0.40 & 0.40 \\
Claim score & 0.905 & 0.774 & 0.853 & 0.801 & 0.543 & 0.679 \\
Answers the brief & B & A & B & A & C & F \\
Supervision & C & A & B & B & C & F \\
\end{tabular}
\end{ruledtabular}
\caption{\label{tab:si:trial_table}
\textbf{The six live stacking-order sessions.}
All values for the six main-text sessions that determined the stacking order of a multilayer graphene device (Figs.~\ref{fig:main:2} and \ref{fig:main:matrix}): HERA's conclusion and its confidence level, agreement with the expert's sealed conclusion, interventions before and after the conclusion, active and wall-clock time, cost at list price, the claim score, and the two expert letter grades, for whether the work answers the brief and for the supervision a further session would need.
Sweeps counts the number of measurements in the session archive before and after HERA's conclusion; the sweeps after are any measurements the operator requested.
Expert agrees is yes when stacking order and confidence level both match the sealed conclusion, order only when the order matches at a lower confidence level, and no when the order differs.
Standby is the time HERA spent waiting on the operator or a provider outage, and sweep time after conclusion the duration of the runs the operator requested after it (Sec.~\ref{sec:si:interventions}).
Claims and items are the counts the expert graded; inference, decision, and verdict are the mean grades by claim type that Fig.~\ref{fig:main:matrix}b plots (Methods).
}
\end{table}

\begin{figure}[htbp]
\centering
\providecommand{\subsz}{\fontsize{6}{7}\selectfont}
\definecolor{concGood}{RGB}{17, 112, 170}
\definecolor{concBad}{RGB}{209, 97, 13}
\definecolor{concWrong}{RGB}{190, 40, 40}
\begin{tikzpicture}[font=\footnotesize,
  chip/.style={rounded corners=1.5pt, minimum width=0.86cm, minimum height=0.32cm, inner sep=0.5pt, font=\tiny\sffamily\bfseries},
  sound/.style={chip, fill=concGood!22, draw=concGood!70, text=concGood!80!black},
  shaky/.style={chip, fill=concBad!22, draw=concBad!80, text=concBad!80!black},
  wrong/.style={chip, fill=concWrong!22, draw=concWrong!80, text=concWrong!85!black},
  na/.style={chip, fill=black!8, draw=black!40, text=black!60},
  cb/.style={line width=1.0pt},
]
\def\pitch{0.98}
\def\txtsound{SOUND}\def\txtshaky{SHAKY}\def\txtwrong{WRONG}\def\txtna{N/A}
\newcommand{\chainunit}[5]{%
  \node[#3, #4] at ({#2*\pitch}, #1) {\csname txt#3\endcsname};
  \node[anchor=north, font=\subsz\sffamily, text=black!65, align=center, inner sep=1pt] at ({#2*\pitch}, #1-0.17) {#5};}
\newcommand{\rowhead}[3]{\node[anchor=east, align=right, font=\footnotesize] at (-0.55, #1) {\textbf{#2}\\[-2pt]{\subsz\sffamily\color{black!60} #3}};}
\def\rB{0}\def\rC{-1.0}\def\rD{-2.0}\def\rE{-3.0}\def\rF{-4.0}

\rowhead{\rB}{device 2 (4L)}{Fable 5 $\cdot$ corpus}
\chainunit{\rB}{0}{sound}{}{gate C}
\chainunit{\rB}{1}{sound}{cb}{gate re}
\chainunit{\rB}{2}{sound}{cb}{gatemap}
\chainunit{\rB}{3}{sound}{cb}{SdH e$^-$}
\chainunit{\rB}{4}{shaky}{cb}{Si test}
\chainunit{\rB}{5}{sound}{cb}{fan}
\chainunit{\rB}{6}{sound}{}{LK}
\chainunit{\rB}{7}{shaky}{cb}{LK v2}
\chainunit{\rB}{8}{sound}{}{close}
\chainunit{\rB}{9}{sound}{cb}{call}

\rowhead{\rC}{device 2 (4L)}{Fable 5 $\cdot$ corpus $\cdot$ repeat}
\chainunit{\rC}{0}{sound}{}{gate C}
\chainunit{\rC}{1}{shaky}{}{gatemap}
\chainunit{\rC}{2}{shaky}{cb}{gmap re}
\chainunit{\rC}{3}{shaky}{}{SdH e$^-$}
\chainunit{\rC}{4}{shaky}{cb}{SdH h$^+$}
\chainunit{\rC}{5}{sound}{}{fan}
\chainunit{\rC}{6}{sound}{}{re-land}
\chainunit{\rC}{7}{sound}{cb}{re-land 2}
\chainunit{\rC}{8}{shaky}{}{LK}
\chainunit{\rC}{9}{sound}{}{close}
\chainunit{\rC}{10}{sound}{cb}{call}

\rowhead{\rD}{device 2 (4L)}{Opus 5 $\cdot$ corpus}
\chainunit{\rD}{0}{sound}{}{read}
\chainunit{\rD}{1}{sound}{}{range}
\chainunit{\rD}{2}{shaky}{}{settling}
\chainunit{\rD}{3}{sound}{}{SNR}
\chainunit{\rD}{4}{shaky}{}{gate C}
\chainunit{\rD}{5}{sound}{}{reset}
\chainunit{\rD}{6}{sound}{}{gatemap}
\chainunit{\rD}{7}{sound}{cb}{cal. fan}
\chainunit{\rD}{8}{sound}{cb}{BG fan}
\chainunit{\rD}{9}{sound}{}{SdH}
\chainunit{\rD}{10}{sound}{}{close}
\chainunit{\rD}{11}{na}{}{state}
\chainunit{\rD}{12}{sound}{cb}{fan $\tau$}
\chainunit{\rD}{13}{sound}{}{close 2}
\chainunit{\rD}{14}{sound}{cb}{call}

\rowhead{\rE}{device 2 (4L)}{Fable 5 $\cdot$ \emph{no corpus}}
\chainunit{\rE}{0}{shaky}{cb}{gate C}
\chainunit{\rE}{1}{wrong}{}{CNP hunt}
\chainunit{\rE}{2}{sound}{}{gatemap}
\chainunit{\rE}{3}{sound}{}{relaunch}
\chainunit{\rE}{4}{sound}{cb}{band map}
\chainunit{\rE}{5}{shaky}{}{fan ramp}
\chainunit{\rE}{6}{wrong}{cb}{fan re}
\chainunit{\rE}{7}{wrong}{cb}{fan 8$\to$0}
\chainunit{\rE}{8}{shaky}{cb}{T + call}
\chainunit{\rE}{9}{sound}{}{close}
\chainunit{\rE}{10}{sound}{}{$(n,D)$}

\rowhead{\rF}{device 2 (4L)}{Opus 5 $\cdot$ \emph{no corpus}}
\chainunit{\rF}{0}{sound}{cb}{CNP}
\chainunit{\rF}{1}{wrong}{cb}{re-range}
\chainunit{\rF}{2}{shaky}{}{gatemap}
\chainunit{\rF}{3}{wrong}{cb}{gmap re}
\chainunit{\rF}{4}{wrong}{cb}{fan}
\chainunit{\rF}{5}{wrong}{cb}{SdH}
\chainunit{\rF}{6}{wrong}{cb}{T + call}
\chainunit{\rF}{7}{sound}{}{close}
\chainunit{\rF}{8}{sound}{}{erratum}
\chainunit{\rF}{9}{sound}{}{close 2}
\chainunit{\rF}{10}{sound}{}{$(n,D)$}

\begin{scope}[shift={(0, 0.55)}]
  \node[sound, minimum width=0.6cm, minimum height=0.26cm] at (0.30,0) {SOUND};
  \node[shaky, minimum width=0.6cm, minimum height=0.26cm] at (1.05,0) {SHAKY};
  \node[wrong, minimum width=0.6cm, minimum height=0.26cm] at (1.80,0) {WRONG};
  \node[na, minimum width=0.6cm, minimum height=0.26cm] at (2.55,0) {N/A};
  \node[sound, cb, minimum width=0.6cm, minimum height=0.26cm] at (3.55,0) {SOUND};
  \node[anchor=west, font=\subsz\sffamily, text=black!65] at (3.90,0) {heavy outline = conclusion-bearing (holds a graded inference or verdict); notebook order, context units omitted};
\end{scope}
\end{tikzpicture}
\caption{\label{fig:si:chain}
\textbf{Session chains of the five tetralayer sessions.}
One strip per session in the order of Fig.~\ref{fig:main:matrix}, every graded notebook unit in reading order with context units omitted, colored by the expert's whole-session ruling on the unit: sound, shaky, wrong, or not applicable.
During the whole-session read, the expert also ruled each notebook unit sound, shaky, or wrong, or marked it not applicable.
A heavy outline marks a unit that holds a graded inference or verdict.
Labels abbreviate the unit's step.
}
\end{figure}
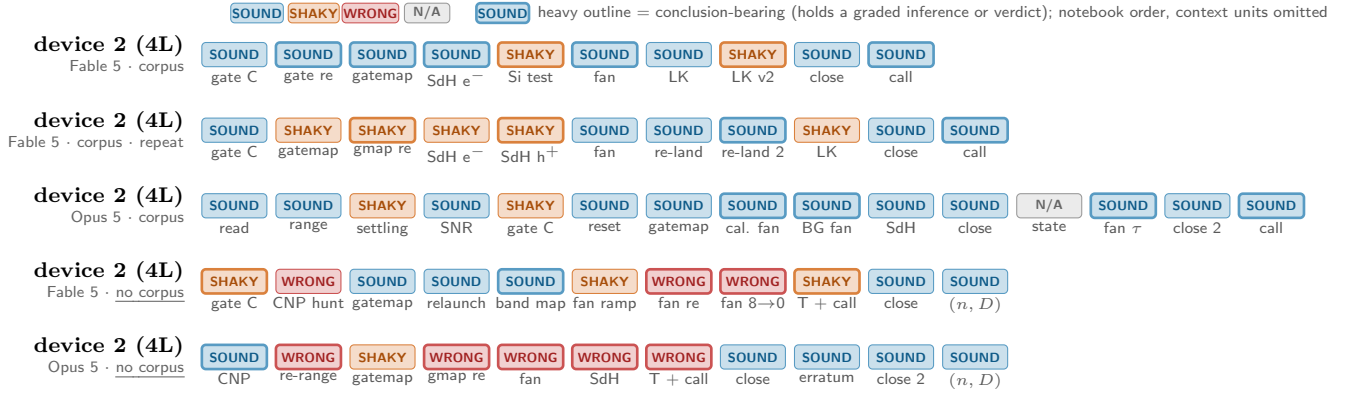

\FloatBarrier

\section{Operator interventions}\label{sec:si:interventions}

On device 1, the operator intervened once before HERA's conclusion.
HERA had stopped itself during its first gate leakage check, and the operator ruled the leakage acceptable.
After the conclusion, the operator requested a finer Landau fan, which resolved the one feature HERA had flagged as unresolved.

On device 2, neither Fable~5 session with the full corpus had any interventions.
Opus~5 with the full corpus had none before its conclusion.
Afterwards, the operator requested a finer Landau fan, which raised HERA's confidence to the expert's, and sent one follow-up message.
Fable~5 with the ablated corpus had none before its conclusion.
During a provider outage, the operator resent the same message six times, and these are not counted.
After its conclusion, the operator asked it to close out the session.
Opus~5 with the ablated corpus had one intervention before its conclusion, when HERA asked for consent to change the lock-in settings.
After its conclusion, the operator sent two corrections during close-out.

Wall-clock time exceeds active time by the close-out, standby, and any measurements the operator requested after the conclusion (Methods).
On device 1, HERA was on standby for 16.6~h.
It waited 11.4~h overnight for the operator's ruling after its safety stop and 5.3~h at close-out.
The fan the operator requested after the conclusion took 5.8~h, followed by a 2.3~h spot check as the field ramped back down.
Opus~5 with the full corpus spent 4.6~h on the fan the operator requested.
Fable~5 with the ablated corpus lost 2.0~h to a provider outage before its conclusion and then waited 8.9~h overnight for the operator's instruction to close out.

\FloatBarrier

\section{Second-expert grading of the claim subset}\label{sec:si:second_expert}

Methods describes how the second expert graded a subset of the claims.
This section compares his grades with the first expert's, by claim type, by session, and item by item.

\begin{table}[htbp]
\small
\begin{ruledtabular}
\begin{tabular}{lccc}
Stratum & Items & Exact agreement & Weighted $\kappa$ \\
\colrule
All & 145 & 0.648 & 0.383 \\
\colrule
Inference claims & 74 & 0.568 & 0.336 \\
Decision claims & 51 & 0.784 & 0.529 \\
Verdict claims & 20 & 0.600 & 0.292 \\
\colrule
Device 1, Fable 5, full corpus & 24 & 0.792 & 0.524 \\
Fable 5, full corpus & 24 & 0.708 & 0.604 \\
Fable 5 repeat, full corpus & 31 & 0.645 & 0.360 \\
Opus 5, full corpus & 19 & 0.632 & 0.497 \\
Fable 5, ablated corpus & 26 & 0.423 & $-0.130$ \\
Opus 5, ablated corpus & 21 & 0.714 & 0.636 \\
\end{tabular}
\end{ruledtabular}
\caption{\label{tab:si:second_expert_agreement}
\textbf{Agreement between the two experts by claim type and by session.}
Items counts the items that neither expert ruled unverifiable, 145 of the 212 in the subset.
$\kappa$ is quadratic-weighted Cohen's $\kappa$ on item values 1, 0.5, and 0.
At the claim level, with a claim's value the mean of the items that expert scored, 53 claims have a value from both experts: exact agreement 0.453, weighted $\kappa$ 0.508.}
\end{table}

\begin{table}[htbp]
\small
\begin{ruledtabular}
\begin{tabular}{lcccc}
Session & Claims in subset & \multicolumn{2}{c}{Claim score on the subset} & Claim score of record \\
& & first expert & second expert & (all claims, first expert) \\
\colrule
Device 1, Fable 5, full corpus & 11 & 0.941 & 0.902 & 0.905 \\
Fable 5, full corpus & 10 & 0.840 & 0.848 & 0.774 \\
Fable 5 repeat, full corpus & 10 & 0.751 & 0.804 & 0.853 \\
Opus 5, full corpus & 8 & 0.781 & 0.824 & 0.801 \\
Fable 5, ablated corpus & 7 & 0.525 & 0.802 & 0.543 \\
Opus 5, ablated corpus & 10 & 0.621 & 0.704 & 0.679 \\
\end{tabular}
\end{ruledtabular}
\caption{\label{tab:si:second_expert_scores}
\textbf{Claim score of each session on the 56-claim subset, under each expert.}
Subset scores are computed as in Methods from each expert's own items over the 53 claims both scored, verdict claims weighted double; the last column is the score of record on the full pool.
Under both experts every session with the full corpus outscores both sessions with the ablated corpus.
Under the second expert the margin is 0.002 (0.804, Fable 5 repeat, full corpus; 0.802, Fable 5, ablated corpus).}
\end{table}

\paragraph{Sealed predictions.}
Before the claims were delivered to the second expert, two predictions were sealed as a SHA-256 hash on 4 September 2026.
The first predicted exact agreement of at least 0.70 between the two experts, and it was not met, at 0.648.
The second predicted that every session with the full corpus would outscore both sessions with the ablated corpus under the second expert's grades, and it was met (Table~\ref{tab:si:second_expert_scores}).

\paragraph{Where the experts disagree.}
The experts disagreed on 51 of the 145 items.
The second expert gave the higher value on 32 of them and the lower on 19.
Nine disagreements were a full step, with one expert ruling the item correct and the other wrong.
Fifteen of the 51 came from Fable~5 with the ablated corpus, which held 26 of the 145 items and was the one session with negative $\kappa$.
In their notes, the first expert lowered grades for poor data before judging the reasoning, while the second graded the reasoning and noted the caveat the agent should have stated.
On one claim from that session, the first expert ruled three items wrong (``fits are terrible \ldots{} is an interpretation of a fit that is wrong \ldots{} CNP did not grow monotonically in field'') and the second ruled them correct.
Even when the two reached the same diagnosis, they could land a step apart.
On a Landau-level fit in the session of Opus~5 with the full corpus, the first expert wrote ``Or your fits are just terrible because you too bad data'' and ruled it wrong, while the second wrote ``I think this dataset is too coarse to do any of the LL fitting'' and ruled it partly correct.
Where the second expert ruled lower and left a note, the note was a physics caveat.
For example, he noted that converting Shubnikov--de~Haas spacing to density ``is not simple'' for multiple orbits on disconnected Fermi surfaces, and that a gap minimum at zero displacement field ``is not a robust criterion'' for stacking order.

\paragraph{Unverifiable rulings.}
The second expert ruled 51 items unverifiable, and the first expert 24, with 8 in common.
The first expert had scored 32 of the second expert's 51 as correct.
By session, the second expert's unverifiable items were 12, 8, 11, 10, 2, and 8, in the order of Table~\ref{tab:si:second_expert_scores}.
On Fable~5 with the ablated corpus, he ruled on nearly every item.
The agreement above is therefore conditional on both experts having ruled.

\FloatBarrier

\section{Superconductivity in the proximitized bilayer graphene device}\label{sec:si:zf1}

\begin{figure}[htbp]
\includegraphics[width=\linewidth]{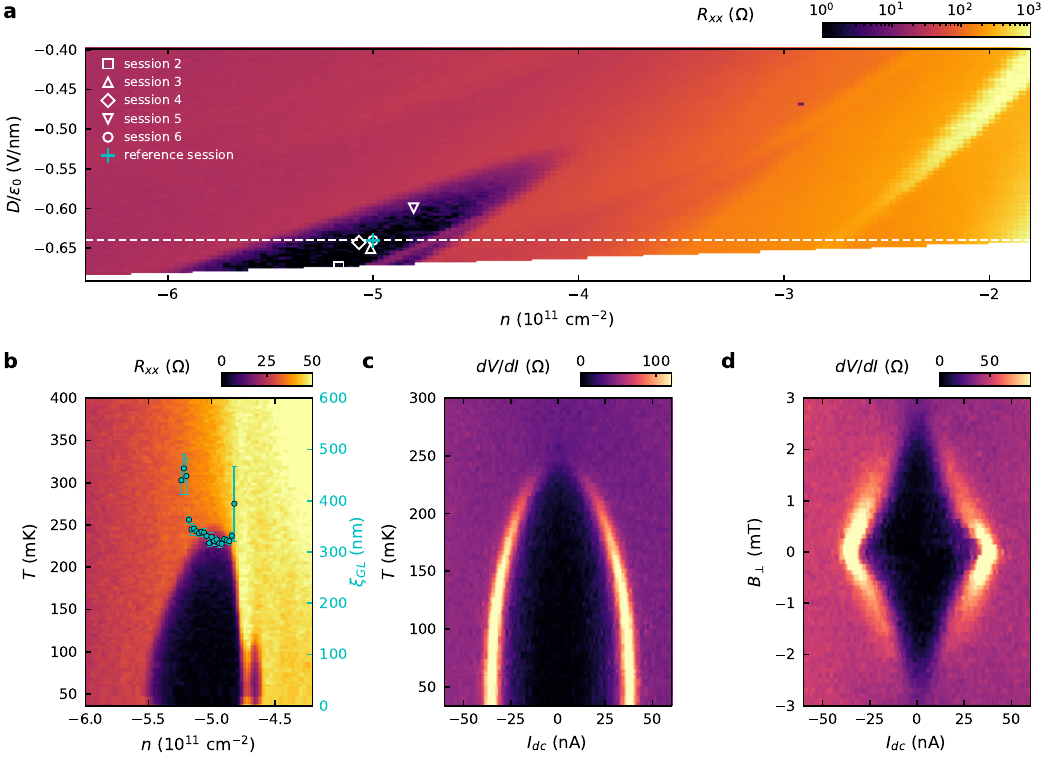}
\caption{\label{fig:si:zf1_characterization}
\textbf{Superconductivity in WSe$_2$-proximitized bilayer graphene.}
(\textbf{a}) Longitudinal resistance $R_{xx}$ as a function of density $n$ and displacement field $D/\epsilon_0$ at 34~mK and effective zero field.
The dashed line denotes $D/\epsilon_0 = -0.64$~V/nm (used in \textbf{b}) and the cyan cross the operating density $n = -5.0\times10^{11}$~cm$^{-2}$ used in \textbf{c} and \textbf{d}.
(\textbf{b}) $R_{xx}$ as a function of $n$ and temperature along the dashed line of (\textbf{a}).
Extracted Ginzburg--Landau coherence length $\xi_{\mathrm{GL}}(n)$ from the 50\,\% criterion of Fig.~\ref{fig:si:zf1_gl} is overlaid on the right axis. 
Error bars correspond to one standard deviation of the fit uncertainty.
(\textbf{c}) Differential resistance $dV/dI$ as a function of d.c.\ bias and temperature at the operating point indicated in \textbf{a}.
(\textbf{d}) $dV/dI$ as a function of bias and perpendicular field at 34~mK; the field axis is relative to the effective zero field at the sample.
All data were acquired by HERA while collaborating with the operator.
}
\end{figure}

This section gives the evidence for phase-coherent transport, the session-by-session results of the search for superconductivity, and the reference characterization of the device.
The recipe lists oscillations of the critical current with perpendicular field among the evidence it seeks.
It asks the agent to attempt this scan, with the field out to the critical field and the bias well past the critical current.
Such oscillations appear only when disorder happens to form a weak link across the device, so their absence does not count against superconductivity.
None has been seen in this device.
Sessions 2 and 4 mapped the differential resistance against bias and perpendicular field at their operating points, out to $\pm 3$ and $\pm 6$~mT, respectively.
Both found a single central lobe with no secondary lobes.
The critical current was largest at the effective zero field and closed by 2.5 to 3~mT.
Session 2 read the lobe as consistent with phase-coherent transport but did not call it an interference pattern.
Session 4 recorded that it had searched for the oscillations and found none.
The other four sessions did not run the scan.
The map of the reference session shows the same single lobe out to $\pm 3$~mT (Fig.~\ref{fig:si:zf1_characterization}d).

\begin{table}[htbp]
\small
\begin{ruledtabular}
\begin{tabular}{lccccclccc}
& & & & \multicolumn{2}{c}{Operating point} & & \multicolumn{3}{c}{Reported} \\
Session & Mode & Hours & Interv. & $n$ & $D/\epsilon_0$ & Call & $I_c$ (nA) & $B_{c\perp}$ (mT) & $T_c$ (mK) \\
\colrule
1 & E0 & 22.7 & 10 & --- & --- & no call & --- & --- & --- \\
2 & E0 & 6.3 & 1 & $-5.17$ & $-0.673$ & not-SC, retracted; SC & 8--11 & 2.5 & 100--200 \\
3 & E0 & 6.6 & 0 & $-5.01$ & $-0.650$ & SC (decisive) & 18--24 & 2.5 & 234 \\
4 & E0 & 9.1 & 3 & $-5.07$ & $-0.643$ & SC (strong) & 30 & 2.5--3 & 300 (onset) \\
5 & E1 & 4.6 & 3 & $-4.80$ & $-0.599$ & consistent-with, hedged & 12--15 & --- & --- \\
6 & E1 & 12.6 & 0 & $-5.00$ & $-0.640$ & SC (decisive) & 28 & 1.8 & 264--290 \\
Ref. & E0 & 87.7 & --- & $-5.00$ & $-0.640$ & reference & 36--37 & 2.2 & 232--236 \\
\end{tabular}
\end{ruledtabular}
\caption{\label{tab:si:zf1_runs}
\textbf{HERA sessions on the WSe$_2$-proximitized bilayer graphene device.}
One row per measurement session, in chronological order.
Mode is the autonomy mode (Methods).
Hours is the wall-clock time of the session and Interv. the number of operator interventions during it.
The operating point ($n$ in $10^{11}$~cm$^{-2}$, $D/\epsilon_0$ in V/nm) is the chosen location where the session performed its characterization of the superconductivity, recomputed from its gate voltages with the reference session's calibration.
Reported values are those HERA returned in its end-of-session conclusion, where SC denotes superconducting.
Session~1 did not commit to a conclusion; session~5 was ended by the operator prior to the temperature ladder.
The reference session was run by HERA in collaboration with the operator and is not scored.}
\end{table}

\begin{figure}[htbp]
\includegraphics[width=\linewidth]{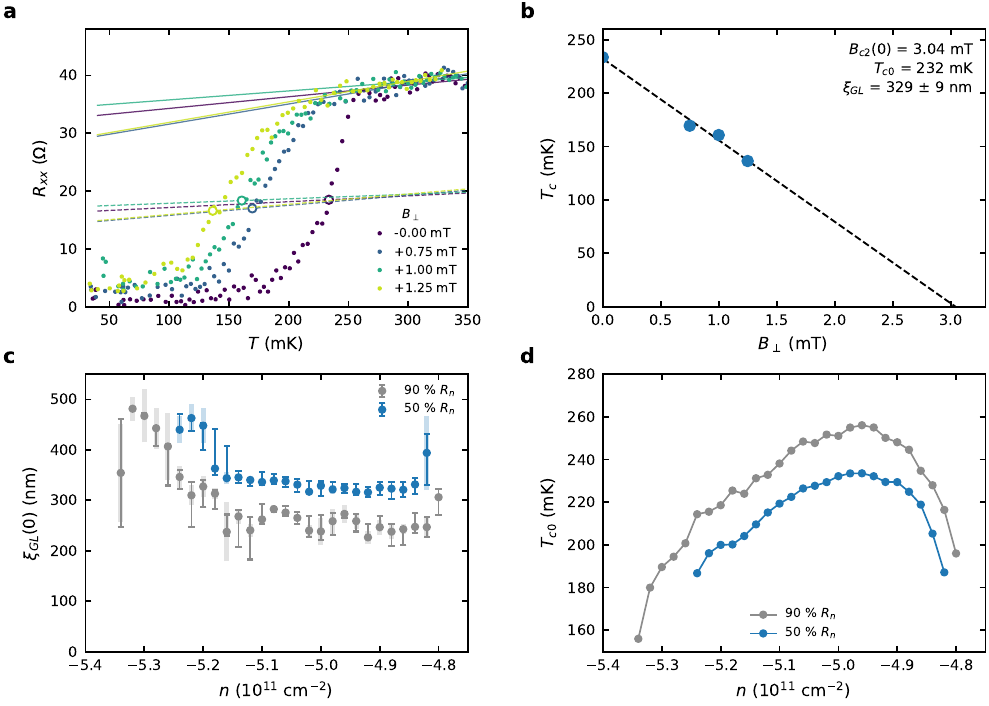}
\caption{\label{fig:si:zf1_gl}
\textbf{Ginzburg--Landau analysis of WSe$_2$-proximitized bilayer graphene.}
(\textbf{a}) $R_{xx}(T)$ at $n = -5.0\times10^{11}$~cm$^{-2}$ and $D/\epsilon_0 = -0.64$~V/nm for perpendicular fields of $0$, $0.75$, $1.00$ and $1.25$~mT.
Solid lines are linear fits to the normal state resistance above 260~mK, dashed lines are half of each fit, and circles mark the transition temperature $T_c$ where the data cross the dashed line.
(\textbf{b}) $T_c$ as extracted from when $R_{xx}(T, B)$ reaches 50\% of the normal state resistance as a function of field with the Ginzburg--Landau fit $T_c/T_{c0} = 1 - 2\pi\xi_{\mathrm{GL}}^2 B/\Phi_0$, giving $B_{c2}(0) = 3.0$~mT and $\xi_{\mathrm{GL}} = 329 \pm 9$~nm.
(\textbf{c}) $\xi_{\mathrm{GL}}$ across the superconducting dome extracted from the 50\,\% criterion of \textbf{b} (blue) and the 90\,\% onset criterion (grey); thin bars span the 40--60\,\% and 85--95\,\% criteria, respectively. 
Faint bars correspond to one standard deviation of the fit uncertainty in the extracted $\xi_{\mathrm{GL}}$.
(\textbf{d}) Zero-field $T_{c0}$ from the same fits.
}
\end{figure}

\begin{figure}[htbp]
\includegraphics[width=\linewidth]{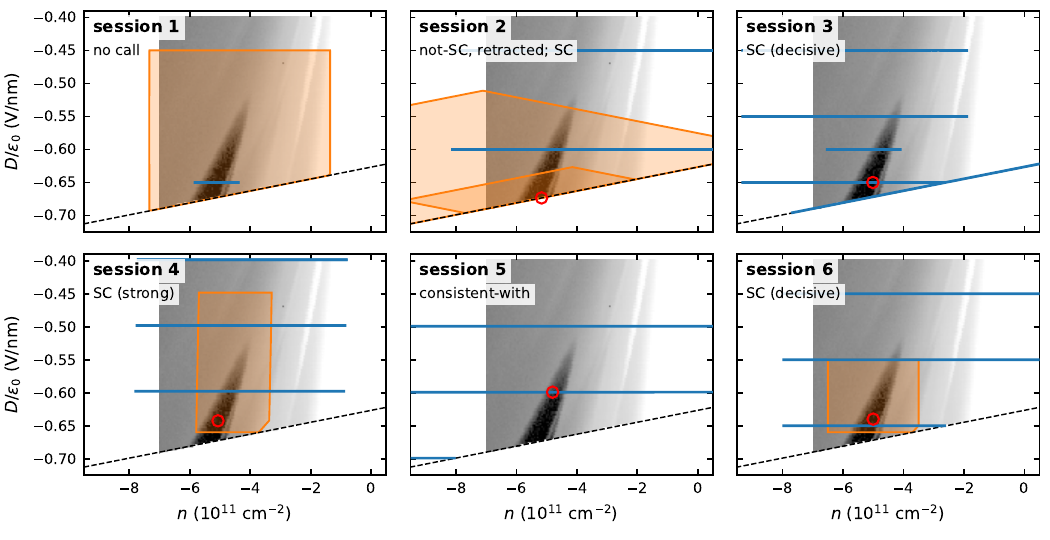}
\caption{\label{fig:si:zf1_coverage}
\textbf{Survey for superconductivity in WSe$_2$-proximitized bilayer graphene.}
Each session's survey for superconductivity is overlaid on the zero-field $R_{xx}(n, D)$ map of the reference session in greyscale.
Every session opened with a low-resolution survey of the full gate window, which is not drawn.
Blue lines denote the line cuts the session ran; orange outlines denote the limits of two-dimensional maps. 
The dashed black line indicates the limit of the accessible gate range. 
The red circle denotes the session's operating point chosen for subsequent characterization.
The label in each panel is the session's conclusion on the presence of superconductivity.
}
\end{figure}

\FloatBarrier

\section{Offline replay: gain at each cutoff}\label{sec:si:replay}

\begin{table}[htbp]
\small
\begin{ruledtabular}
\begin{tabular}{lcccc}
Model & 6~h & 8~h & 12~h & No cutoff \\
\colrule
Fable 5 & $+0.24$ & $+0.10$ & $-0.02$ & $+0.24$ \\
Opus 5 & $+0.26$ & $+0.31$ & $+0.35$ & $+0.35$ \\
Fable 5.1 & $+0.56$ & $+0.72$ & $+0.73$ & $+0.73$ \\
Opus 5.5 & $+0.54$ & $+0.18$ & $+0.13$ & $+0.13$ \\
GPT-6 Astra & $+0.50$ & $+0.46$ & $+0.10$ & $+0.06$ \\
GPT-6 Sol & $+0.81$ & $+0.81$ & $+0.84$ & $+0.84$ \\
\colrule
Pooled & $+0.49$ & $+0.43$ & $+0.35$ & $+0.39$ \\
\end{tabular}
\end{ruledtabular}
\caption{\label{tab:si:budget}
\textbf{Gain in $P(\text{call})$ with the full corpus at each pre-registered cutoff.}
Each entry is the model's mean $P(\text{call})$ over its full-corpus fans minus the mean over its ablated-corpus fans, with a fan planned to run longer than the cutoff counted as zero.
The eight-hour cutoff is the primary one, and the others are pre-registered sensitivity analyses (Methods).
The pooled gain is the mean over the six models, and each pooled gain has an unadjusted permutation $p = 0.0001$.
Between the pre-registered cutoffs, Fable~5's gain is $+0.19$ at seven hours and $-0.13$ at nine.}
\end{table}

\FloatBarrier

\section{Development of the stacking investigation}\label{sec:si:revisions}

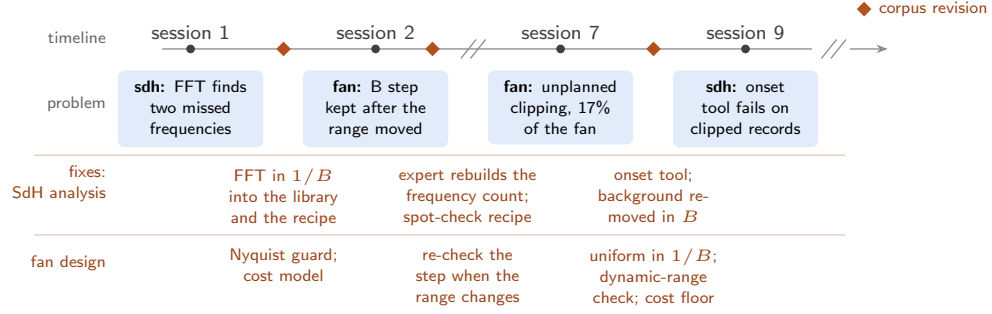
\begin{figure}[htbp]
\centering
\resizebox{0.75\textwidth}{!}{\providecommand{\subsz}{\fontsize{6}{7}\selectfont}
\providecommand{\chipsz}{\fontsize{5.5}{6.3}\selectfont}
\begin{tikzpicture}[font=\scriptsize, x=1cm, y=1cm, >={Stealth[length=1.2mm]},
  burn/.style={draw=none, rounded corners=2pt, fill=heraAgent, align=center, inner sep=3pt, text width=1.32cm, minimum height=0.78cm, font=\chipsz\sffamily, anchor=north},
  fix/.style={anchor=north, align=center, inner sep=1pt, font=\chipsz\sffamily, text=heraLoop!85!black, text width=1.7cm},
  rowlab/.style={anchor=east, align=right, inner sep=1pt, font=\chipsz\sffamily, text=black!60},
  sesslab/.style={anchor=south, inner sep=1.5pt, font=\subsz\sffamily, text=black!75},
]
\def\xa{0.8}\def\xb{2.75}\def\xd{4.7}\def\xe{6.65}
\def\xL{-0.05}
\def\yT{0}        %
\def\yB{-0.22}    %
\draw[black!45, thin] (\xa-0.3, \yT) -- (\xe+0.3, \yT);
\draw[black!45, thin] (\xe+0.3, \yT) -- (7.35, \yT);
\draw[black!45, thin, ->] (7.75, \yT) -- (8.15, \yT);
\draw[black!45, thin] (3.65, \yT-0.1) -- (3.81, \yT+0.1);  \draw[black!45, thin] (3.75, \yT-0.1) -- (3.91, \yT+0.1);
\draw[black!45, thin] (7.45, \yT-0.1) -- (7.61, \yT+0.1);  \draw[black!45, thin] (7.55, \yT-0.1) -- (7.71, \yT+0.1);
\foreach \x/\l in {\xa/1, \xb/2, \xd/7, \xe/9} {
  \fill[black!75] (\x, \yT) circle (1.3pt);
  \node[sesslab] at (\x, \yT+0.04) {session \l};}
\foreach \x in {1.78, 3.35, 5.68}
  \fill[heraLoop] (\x, \yT) ++(-2.2pt,0) -- ++(2.2pt,2.2pt) -- ++(2.2pt,-2.2pt) -- ++(-2.2pt,-2.2pt) -- cycle;
\node[rowlab] at (\xL, \yT+0.12) {timeline};
\fill[heraLoop] (7.9, \yT+0.42) ++(-2.2pt,0) -- ++(2.2pt,2.2pt) -- ++(2.2pt,-2.2pt) -- ++(-2.2pt,-2.2pt) -- cycle;
\node[anchor=west, font=\chipsz\sffamily, text=heraLoop!85!black, inner sep=2pt] at (7.98, \yT+0.42) {corpus revision};
\node[rowlab] at (\xL, \yB-0.36) {problem};
\node[burn] at (\xa, \yB) {\textbf{sdh:} FFT finds two missed frequencies};
\node[burn] at (\xb, \yB) {\textbf{fan:} B step kept after the range moved};
\node[burn] at (\xd, \yB) {\textbf{fan:} unplanned clipping, 17\% of the fan};
\node[burn] at (\xe, \yB) {\textbf{sdh:} onset tool fails on clipped records};
\def\yFa{-1.22}\def\yFb{-2.06}
\draw[heraLoop!30, thin] (\xL-0.8, \yFa+0.1) -- (7.7, \yFa+0.1);
\node[rowlab, text=heraLoop!85!black] at (\xL, \yFa-0.2) {fixes:\\SdH analysis};
\node[fix] at (1.78, \yFa) {FFT in $1/B$ into the library and the recipe};
\node[fix] at (3.72, \yFa) {expert rebuilds the frequency count; spot-check recipe};
\node[fix] at (5.68, \yFa) {onset tool; background removed in $B$};
\draw[heraLoop!30, thin] (\xL-0.8, \yFb+0.1) -- (7.7, \yFb+0.1);
\node[rowlab, text=heraLoop!85!black] at (\xL, \yFb-0.2) {fan design};
\node[fix] at (1.78, \yFb) {Nyquist guard; cost model};
\node[fix] at (3.72, \yFb) {re-check the step when the range changes};
\node[fix] at (5.68, \yFb) {uniform in $1/B$; dynamic-range check; cost floor};
\end{tikzpicture}}
\caption{\label{fig:si:ladder}
\textbf{The development campaign on device 3.}
HERA ran nine sessions on device 3, a trilayer, between 10 and 22 June 2026, and four are shown, numbered by run.
Run 6, a false start, is not counted.
Beneath each session is the problem it exposed, in either SdH analysis (sdh) or fan design (fan), and the two rows below hold the fixes that followed.
A diamond marks a revision of the corpus by the review agent (Methods), and the fixes shown are a sample.
The onset tool also failed in sessions 8 and 10, and it was rewritten in July, after the campaign.
Sessions 1 and 2 ran in E0 (Methods), with Fable~5 for session 1 and Opus~4.8 after.
The last three sessions ran the same brief with no revision of the corpus between them.
Their conclusions varied between ABA and ABC, because the light ABA band lies below resolution at 1.2~K and 8~T on this device.
}
\end{figure}

\end{document}